\documentclass[symmetry,review,accept,moreauthors,pdftex]{Definitions/mdpi} 

\firstpage{1}
\pubvolume{17}
\issuenum{6}
\articlenumber{974}
\pubyear{2025}
\copyrightyear{2025}
\externaleditor{Academic Editor: {Theodore E. Simos}} 
\datereceived{9 May 2025}
\daterevised{6 June 2025}
\dateaccepted{10 June 2025}
\datepublished{19 June 2025}
\hreflink{\textls[-35]{https://doi.org/10.3390/sym17060974}} 

\usepackage{amsfonts,amsmath,amssymb,amsthm,mathtools,xcolor,booktabs,graphicx,array,colordvi,braket,slashed}

\def\Mp{M_{\textrm{Pl}}}
\def\Lp{\ell_{\textrm{Pl}}}
\def\Ep{E_{\textrm{Pl}}}
\def\Elvn{E_{\textrm{LV},n}}
\def\Elv{E_{\textrm{LV}}}
\def\Elvg{E_{\textrm{LV}\gamma}}
\def\Elvnu{E_{\textrm{LV}\nu}}
\def\Elve{E_{\textrm{LV}e}}
\def\Ms{M_{s}}
\def\Ls{\ell_{s}}
\def\gs{g_{s}}
\def\Dp{{\cal D}}
\def\Md{M_{\Dp}}
\def\S{{\cal S}}
\def\C{{\cal C}}
\def\Q{{\cal Q}}
\def\G{{\cal G}}

\def\U{{\cal U}}
\def\bfV{\mbox{\boldmath$\U$\unboldmath}}
\def\V{{\cal V}}
\def\bfV{\mbox{\boldmath$\V$\unboldmath}}
\def\pd{\partial}
\def\sd{\textrm{d}}
\def\e{\textrm{e}}
\def\I{\textrm{i}}
\def\Or{{\cal O}}
\def\D{\Delta}
\def\pv{{\bf p}}
\def\pn{\lvert\pv\rvert}
\def\kv{{\bf k}}
\def\kn{\lvert\kv\rvert}
\def\sdla{\langle\!\langle}
\def\sdra{\rangle\!\rangle}
\def\Bdla{\Bigl\langle\!\!\Bigl\langle}
\def\Bdra{\Bigr\rangle\!\!\Bigr\rangle}
\def\alp{\alpha^{\prime}}
\def\fd{\mathfrak{d}}
\def\K{{\cal K}}
\def\z{\mathfrak{z}}
\def\Y{{\cal Y}}
\def\be{\begin{equation}}
\def\la{\label}
\def\ee{\end{equation}}
\def\bi{\begin{itemize}}
\def\im{\item}
\def\ei{\end{itemize}}

\Title{Probes for String-Inspired Foam, Lorentz, and CPT Violations in Astrophysics}

\TitleCitation{Probes for String-Inspired Foam, Lorentz, and CPT Violations in Astrophysics}

\Author{Chengyi Li $^{1}$\orcidA{} and Bo-Qiang Ma $^{1,2,3,*}$\orcidB{}}

\AuthorNames{Chengyi Li and Bo-Qiang Ma}

\AuthorCitation{Li, C.; Ma, B.-Q.}

\address{%
$^{1}$ \quad School of Physics, Peking University, Beijing 100871, China; lichengyi@pku.edu.cn~(C.L.); mabq@pku.edu.cn~(B.-Q.M.)\\
$^{2}$ \quad School of Physics, Zhengzhou University, Zhengzhou 450001, China\\
$^{3}$ \quad Center for High Energy Physics, Peking University, Beijing 100871, China}

\corres{Correspondence: mabq@pku.edu.cn}

\abstract{Lorentz invariance is such a basic principle in fundamental physics that it must be constantly tested and that any proposal of its violation and breakdown of CPT symmetry, that might characterize some approaches to quantum gravity, should be treated with care. In this review we examine, among other scenarios, such instances in supercritical~(Liouville) string theory, particularly in some brane models for ``quantum foam''. Using the phenomenological formalism introduced here, we analyze the observational hints of Lorentz violation in time-of-flight lags of cosmic photons and neutrinos which fit excellently stringy space-time foam scenarios. We further demonstrate how stringent constraints from other astrophysical data, including the recent first detections of multi-TeV events in $\gamma$-ray burst 221009A and PeV cosmic photons by the Large High Altitude Air Shower Observatory~(LHAASO), are satisfied in this context. Such models thus provide a unified framework for all currently observed phenomenologies of space-time symmetry breaking at Planckian scales.}


\begin{document}
\section{Introduction and Summary}\label{sec:1}

To this day, particle-physics theory and the respective phenomenology enjoy a Lorentz-symmetric formalism dictated by special relativity~(SR). As a consistent gauge field theory model of electromagnetic, weak and strong interactions, the standard model~(SM) of particle physics is by construction Lorentz invariant, respecting CPT symmetry as a result. However recent years have witnessed an intense---and ongoing---interest in potential tiny deviations from these fundamental symmetries, especially the exact local Lorentz invariance of general relativity~(GR) that was attributed to Einstein to interpret gravity as space-time curvature. On the theoretical side, ideas stemming from quantum gravity~(QG) which primarily aims at conquering the ultraviolet~(UV) divergence of quantum GR led to speculation that these may not be exact for QG~\cite{Shao:2010wk,Amelino-Camelia:2008aez,Ma:2012dv}, albeit they survived many stringent tests~\cite{Coleman:1998ti}, even the ones making use of high-energy particle probes. On the observational side, with the increased sensitivity of experiments, especially cosmic multimessenger ones, to the effects of violation of Lorentz~(LV) and/or, CPT symmetry~(CPTV), Fermi measurements~\cite{Fermi-LAT:2009owx,FermiGBMLAT:2009nfe} of time delays in the arrivals of high-energy $\gamma$-ray burst~(GRB) photons as well as IceCube observations~\cite{IceCube:2013cdw,IceCube:2013low,IceCube:2014stg} for TeV--PeV neutrinos~(some of which appear not incompatible with GRBs~\cite{Amelino-Camelia:2015nqa,Amelino-Camelia:2016fuh,Amelino-Camelia:2016wpo,Amelino-Camelia:2016ohi,Amelino-Camelia:2017zva,Huang:2018ham,Huang:2019etr,Huang:2022xto}), played an indispensable role in searching for such QG footprints.~(See, e.g.,~\cite{He:2022gyk} for a review of one series of studies~\cite{Shao:2009bv,Zhang:2014wpb,Xu:2016zxi,Xu:2016zsa,Amelino-Camelia:2016ohi,Amelino-Camelia:2017zva,Xu:2018ien,Liu:2018qrg,Chen:2019avc,Zhu:2021pml,Zhu:2021wtw,Zhu:2022usw,Song:2024and,Song:2025qej,Song:2025apj,Song:2025akr} in this direction.)

Although conventional astrophysical source effects, which currently are far from being understood, might be the dominant reason for the retarded photon arrivals as observed in Fermi GRBs, the data from Fermi, and/or other photon-dedicated telescopes can be readily adopted to test LV models. Even the most generic models predict naturally such shifts in the travel times of~(ultrarelativistic) cosmic particles given the conjecture that space-time, probed at a small enough scale, could appear complicated---something akin in complexity to a turbulent froth that John Wheeler~\cite{Wheeler:1964qna,Wheeler:1998vs} dubbed \emph{space-time foam}. If space-time itself, like all matter and energy, indeed undergoes quantum fluctuation~(on time scales $\sim 1/\Mp$, with $\Mp\simeq 1.22\times 10^{19}$~GeV being the Planck mass), our common perception of relativistic, continuous space-times, like Lorentz invariance, may be sacrificed on the altar of QG. There will be an \emph{energy-dependent} variation in the speed of an energetic photon or neutrino \emph{in vacuo}, $\delta\coloneqq (v-c)/c$~\cite{Amelino-Camelia:1997ieq,Jacob:2006gn,Addazi:2021xuf}, leading to flight-time differences~\cite{Jacob:2008bw} between energetic particle~(of energy $E$) and the low-energy light which travels at the light speed of SR, $c$~(hereafter, we use natural units, where $c=\hslash=1$). On rather general grounds~\cite{Li:2022sgs},
\be\la{eq:mdr}
\delta(E)=-\frac{n+1}{2}s_{n}\Bigl(\frac{E}{\Elvn}\Bigr)^{n},\qquad E\ll\Elvn,
\ee
where $s_{n}=+1$~($-1$), the sign factor, refers to the \emph{subluminal}~($\delta<0$)~(\emph{superluminal}~($\delta>0$)) scenario, while $\Elvn$ is the LV energy scale, the scale at which $n$th-order LV sets in, expected to be of the order of the Planck energy~($\Ep$), and to be determined experimentally. Modified dispersion relations~(MDRs), implied by Equation~(\ref{eq:mdr}), as compared to the one in the normal vacuum, $E^{2}=\pv^{2}+m^{2}$ for a particle of mass $m$, have two major classes of consequences that can be tested in astrophysical contexts:
\bi
\im[(i)] Energy-dependent lag in the propagation of massless~(or, almost massless) signal from cosmically remote source to Earth, as mentioned;
\im[(ii)] Birefringence~(i.e., different speeds among helicities), and MDR-modified~(or novel) threshold reactions based on the assumption of standard energy--momentum conservations~(owing to a lack of any particular prior knowledge).
\ei
The second constitutes an independent portion of the entire effort on constraining Lorentz violation, providing some of the tightest restrictions for models of new physics, a topic that we shall cover in this review.

Due to the high precision of observations, many Lorentz tests were motivated, yielding fruitful results and, the potentiality that certain QG scenarios could account for all or part of the available phenomenologies raised a vibrant interest. It is known that any LV model able to reproduce the measured time delays of several GRB photons while admitting an effective-field-theory~(EFT) formulation is in tension with other astrophysical measurements~\cite{Li:2021tcw}~(see also~\cite{Jacobson:2004rj,Mattingly:2005re,Jacobson:2005bg,Liberati:2009pf,Liberati:2013xla}). However, following~\cite{Mavromatos:2010pk} and the recent suggestion of~\cite{Li:2021gah,Li:2021eza}, a QG framework beyond local EFTs, capable of accommodating data on time-of-flight of cosmic photons~(and neutrinos) while successfully circumventing other observational restrictions does exist and has a string/D(irichlet)-brane theory origin. In particular in~\cite{Li:2023wlo,Li:2022sgs} and~\cite{Li:2024crc}, it is not only implied that the theory could possibly be in accordance with either a CPT-violating neutrino propagation suggested from previous analyses of~\cite{Amelino-Camelia:2016ohi,Amelino-Camelia:2017zva,Huang:2018ham,Huang:2019etr,Huang:2022xto}, or a necessarily subluminous one, a possibility recently enhanced by a new study~\cite{Amelino-Camelia:2022pja} using revised data~\cite{IceCube:2020wum}, but shown that it might also explain, in case of breaking CPT between neutrinos~($\nu$'s) and antineutrinos~($\bar{\nu}$'s), the quasi-stability against superluminal $\bar{\nu}$-decays in the corresponding string vacuum. In this review, we present a summary for these findings and prospects for probing microscopic space-time symmetries dictated by such models.

In Section~\ref{sec:2} we follow the historical path, mentioning some of the early results of space-time fluctuations from QG analysts. For the purposes of the review, in Section~\ref{sec:3} we focus on instances in string theory where Lorentz symmetry may be broken. In particular, we explain how such LV may arise in a generic noncritical~(Liouville) inspired context of strings~\cite{Ellis:1992eh,Ellis:1998dw}, while in Section~\ref{sec:4} we discuss an example of the latter, that is the above-mentioned D-brane approach to \emph{foamy} space-time, whose nontrivial properties are of the type that is compatible with stringy gauge symmetries, and thus primarily affect the propagation of photons and neutrinos. The discussion will include three, different scenarios which are known to induce distinguishable consequences for particle's kinematics. By adopting the phenomenological approach to LV, c.f.~(\ref{eq:mdr}), in Section~\ref{sec:5} we summarize current observational hints of LV energy-dependent time lags for cosmic photons and neutrinos; in Section~\ref{sec:6}, we briefly review some complementary constraints due to the lack of signatures of other LV effects in astrophysical data. Finally Section~\ref{sec:7} states our conclusions.

It should be stressed that the review is from a personal perspective and is by no means complete. We do hope, however, that it captures some essential features of this rapidly expanding field with some recent progress. In this respect we include several latest results from the surprising discoveries of LHAASO~\cite{LHAASO:2019qtb,Cao:2021vqs} on PeV-energy photons~\cite{LHAASO:2021gok,LHAASO:2021cbz,LHAASO:2023uhj} and on the extraordinary, brightest-of-all-time GRB~221009A~\cite{Huang:2022grb,LHAASO:2023kyg,LHAASO:2023lkv}, whose sensitivities to certain types of LV reach or even surpass Planck-scale sensitivity under some circumstances~\cite{LHAASO:2021opi,Chen:2021hen,Li:2022ugz,He:2022jdl,He:2023ydr} although, on the other side~\cite{Li:2021duv,Li:2022vgq,Li:2022wxc,Finke:2022swf,Li:2023rgc,Li:2023rhj} they appear to support some kind of LV effects. We shall put the emphasis on the connections of these experimental results to the aforementioned D-brane/string scenarios, and attempt to answer the question as to whether there is the possibility of attributing the current LV phenomenologies, at least partly, to the~(stringy) QG effects such models entail.

\section{Space-Time Symmetry and Quantum Gravity}\label{sec:2}

Through more than a century of experimental tests with high precision, Einstein's~(classical) theories of relativity that entail Lorentz invariance, at least locally, now seem to stand fully verified. However, as one of the most ambitious goals of modern physics, the quest for a quantum theory of gravity led to suspect that there could be a failure of this symmetry. Indeed, analysts of quantum gravity expect that the fabric of space-time should fluctuate over distance scales inversely proportional to the Planck scale, $\Mp$. It was on this conceptual basis that the conventional special-relativistic relation between momentum and energy might get modified, reflecting a breakdown of Lorentz symmetry~\cite{Shao:2010wk,Amelino-Camelia:2008aez,Ma:2012dv,Addazi:2021xuf,He:2022gyk}. It is tempting to go one step forward and consider the possibility, that discrete symmetry of space-time continuum, like the combination of parity, charge conjugation, and time-reversal may not be a true feature of QG. Although CPT symmetry arises from some of the basic constituents of local, point-particle quantum field theory~(QFT), nonetheless in QG realm, the concept of a \emph{local effective QFT} Lagrangian may \emph{break down}.

Albeit the goal of a theory capable of incorporating quantum theory and GR remains unattainable, various ways of tackling certain aspects of the problem have been suggested, including canonical gravity~\cite{Wheeler:1964qna}, Liouville noncritical strings~\cite{Ellis:1992eh,Ellis:1998dw} and D-branes~\cite{Polchinski:1995mt,Polchinski:1996na,Polchinski:1998rr}. Within the latter framework, to be discuss in greater detail later on, the situation resembles that of a quantum-decoherent motion of matter in open dynamical systems interacting with an environment~\cite{Hawking:1982dj,Ellis:1983jz}. There are other fascinating schemes for QG; only to mention a few, asymptotically safe gravity, causal sets, modified, and nonlocal gravities. Admittedly, not all proposals are at variance with standard space-time symmetries. For example in causal sets~\cite{Dowker:2003hb} contrary to what was often thought, a fundamental space-time discreteness needs not contradict Lorentz symmetry.

Nonetheless, specific hints of Lorentz violation have come from tentative calculations in the two approaches to quantum gravity proper, that perhaps, have attracted more attention, that is, the loop-variable approach~(formerly termed canonical QG) and~(super)string theory/cosmology:

\bi
\im It has been established that, in attempting to quantize Einstein's theory of GR, in the familiar framework of local 4-dimensional field theories and~(nontrivial, topological) extensions thereof, i.e., using a canonical formalism, like the loop programme~\cite{Ashtekar:2021kfp}, where the states of the theory are described by functions of ``spin networks'', a ``polymer-like'' discrete structure of space-time emerges. In the semiclassical regime of such theories, the gravitational degrees of freedom are in a ``weavy state'' $\Ket\varpi$~\cite{Ashtekar:1992tm}, characterized by a length scale $\ell_{\varpi}\gg\Lp$~(here $\Lp$ is the Planck length) such that, although macroscopically the~(emergent) space-time looks Minkowski, nevertheless novel features emerge at scales smaller than the $\ell_{\varpi}$ where loop QG effects set in:
\be\la{eq:lws}
\Bra\varpi g_{\mu\nu}\Ket\varpi=\eta_{\mu\nu}+\Or(\Lp/\ell_{\varpi}).
\ee
Equations of motion for the propagating particles~\cite{Gambini:1998it,{Alfaro:1999wd,Alfaro:2002xz},Alfaro:2001rb} are modified in a Lorentz-noninvariant way in such states, as becomes clear upon rewriting~(\ref{eq:lws}), on dimensional grounds, as $\Bra\varpi g_{\mu\nu}\Ket\varpi=\eta_{\mu\nu}+\Or(\ell_{\varpi}E)$, with $E$ being the energy of the particle probe. Such effects then, lead to nontrivial MDRs, similar in nature to the one~(\ref{eq:mdr}) postulated in the generic phenomenological approach. However the explicit form of such dispersion relations depends highly on the details of the model. In models of~\cite{Gambini:1998it,Alfaro:2001rb} superluminal propagation of matter probes is allowed, implying birefringence effects, while it may happen~\cite{Li:2022szn} that such superluminal signals of light are eliminated, for some reasonable choices of the model parameters, in which case only subluminal photons are present but for neutrinos, a dependence on their helicities is implied. It has been noted that, in this latter scenario of loop gravity, there would be compatibility~\cite{Li:2022szn} with the hints of LV time lag discussed later on in this paper.
\im An analogous effect may be anticipated in string theory and string cosmology, where gravity is included from the outset, particularly when one considers the nonperturbative formulation of strings, which is known as M-theory. This involves solitonic states such as D-branes~\cite{Polchinski:1995mt,Polchinski:1998rr}, including ``D-particle''~(i.e., D0-brane) defects in space-time. We first note that the vacuum considered in conventional, ``old'' string theory, living in a \emph{critical} dimension of~(flat) target space-time, exhibits essentially \emph{no} Lorentz violation. In such a case there is a basic symmetry, dubbed conformal invariance, which allows a consistent path-integral formulation of the $\sigma$-model describing the motion of strings from a first quantization viewpoint~\cite{Green:2012oqa,Green:2012pqa}, and which restricts the background fields to their target-space classical equations of motion, thereby offering an important link between consistent worldsheet quantum geometry with target-space dynamics. It is this symmetry that guarantees Lorentz invariance in critical dimensions of space-times, and the standard dispersion relations for stringy excitations.\vspace{0.05cm}

However, critical strings, and the corresponding conformal field theories~(CFTs), describe only \emph{fixed} backgrounds, $g_{\ast}$~(i.e., \emph{equilibrium} situations, in light of the conformal $\sigma$-model actions $\S_{0}[g_{\ast}]$, whose central charge, $\C$, has the critical value $c^{\ast}$), thus inadequate for the needs of investigating stochastic space-time foam backgrounds~(if one adopts the point of view that they exist), recoil effects of stringy D-brane solitons, and in general situations that involve a \emph{change} of the background~(field) over which the string propagates. Perhaps two approaches that have been invoked to tackle the issue have been string field theory, a proposal of second quantization of string theory, which remains poorly understood till this day, on the one hand and the \emph{noncritical}~(Liouville) string theory on the other.
\bi
\im[$\divideontimes$] \underline{String Noncriticality}~\cite{David:1988hj,Distler:1988jt,Distler:1989nt,Mavromatos:1989nf}\vspace{0.15cm}

Formally, space-time foam situations, of our interest, involving singular quantum fluctuations at microscopic scales imply out-of-equilibrium processes, associated in a~(perturbative) first-quantized string, to \emph{nonconformal} couplings/background-fields, $g^{I}$. The \emph{noncriticality} nature of quantum-foam vacua refers to the fact that, there are central charge deficits, $\Q^{2}\propto\C[g]-c^{\ast}$. In such a case, the corresponding $\sigma$-model requires ``dressing'' with the Liouville mode $\phi$, a fully fledged dynamical field, whose presence is essential in restoring the lost conformal invariance. Although this is done at the cost of having an extra target-space dimension, which is timelike if the string is \emph{supercritical}, i.e., $\Q^{2}>0$~\cite{Antoniadis:1988aa,Antoniadis:1988vi,Antoniadis:1990uu,Ellis:1995nv}, $\phi$ may be viewed as a local~(irreversible) renormalization group~(RG) scale $\mu$ and hence its worldsheet zero mode is \emph{identified} with target time $t$~\cite{Ellis:1992eh}. The condition implying the recovery of conformal symmetry for the dressed $g^{I}(\phi\rightarrow t)$ reads
\be\la{eq:gcyc}
\ddot{g}^{I}+\Q(t)\dot{g}^{I}=-\textrm{sgn}(\C[g,t]-c^{\ast})\beta^{I}(g(\phi))+\ldots,
\ee
where the $\ldots$ stand for corrections that are of importance if one goes further away from equilibrium, the dot denotes temporal~(Liouville) derivative, and $\beta^{I}$ are the relevant RG $\beta$-functions, or the so-called Weyl anomaly coefficients, proportional to the \emph{off-shell} variations of a target-space effective action $\S_{\textrm{eff}}$ for the string theory at hand~\cite{Osborn:1989td}:
\be\la{eq:bgf}
\beta^{I}=\G^{IJ}\frac{\delta\S_{\textrm{eff}}}{\delta g^{J}}[g(\mu)],
\ee
where $\G^{IJ}$ is the inverse of the Zamolodchikov metric~\cite{Zamolodchikov:1986gt} in the space of string ``theories''~(backgrounds) $\{g^{I}\}$. The latter, the off-shell $\beta$-functions vanish if stringy $\sigma$-models are at their conformal~(fixed) points in $g^{I}$ space, thereby, corresponding to the usual critical equilibrium string. From the cosmological point of view, equations of the form~(\ref{eq:gcyc}) characterize Liouville nonequilibrium $\Q$-cosmologies~\cite{Ellis:2005qa,Mavromatos:2009pm} replacing the standard Einstein equations.
\ei
We describe in the next section how Lorentz invariance properties of the string vacuum may be broken in this framework, in the sense of leading to nontrivial \emph{refractive} indices, or MDRs, for photons \emph{in vacuo}. To our knowledge, this was the first instance where such effects have been explored in concrete models of QG. Later, the above loop and many other approaches to MDRs were put forward, for a variety of reasons we do not mention here.
\ei
Clearly, quantum gravity is still a programme far from completion, and, it seems to be a field like no other before, an area without much experimental guidance~\cite{Carlip:2001wq}. However, as we indicate below, likely ``relic'' effects of QG-motivated space-time symmetry violations~\cite{Jacobson:2004rj,Mattingly:2005re,Jacobson:2005bg,Liberati:2009pf,Shao:2010wk,Liberati:2013xla} propagate down to low-energy~(relative to the Planck energy) regime, where they can be probed. In a sense, they have already provided us a valuable opportunity to experimentally falsify the predictions of the underlying new-physics, particularly QG theories, proposed so far. For this reason, these observational consequences that LV may induce have served as what was often termed ``\emph{windows on quantum gravity}''. Indeed, even small corrections, in principle, have a chance to be magnified into measurable quantities, especially when one deals with the extreme environments, like situations encountered in astroparticle physics. In this respect they have led already to a rich phenomenology of LV~\cite{Addazi:2021xuf}, that is the focus of the present paper. It is for the same reason why much activity in this line is under way~\cite{AlvesBatista:2023wqm}. It will hopefully redirect the study of QG on a healthier track once we learn to recognize physically relevant facts about such effects.

\section{Models of String-Driven Lorentz Noninvariance}\label{sec:3}

A theory of quantum gravity must describe physics at Planck length, $\Lp\sim 10^{-35}$~m. At such small scales space-time as mentioned before, might be discrete, noncommutative~\cite{Amelino-Camelia:1999jfz}, etc. Therefore the form of Lorentz invariance might be different from the familiar one in low energy domain~(for instance it might be nonlinearly realized at such scales). As we noted, in principle, to solve the problem of~(non)renormalizability of the gravity, one may follow two different ways introduced in the previous section: either abandon perturbative techniques leading to the construction of loop QG, or introduce nonlocality, and/or supersymmetry, as applied to superstrings. Actually, the simplest way to construct renormalizable gravity theory is based on introducing higher derivatives, however to avoid arising ghosts, one has to introduce the Ho\v rava--Lifshitz theory~\cite{Horava:2009uw,Horava:2008jf,Chen:2009ka,Xiao:2009xe}. It constitutes an explicit realization of modified gravity as an UV complete QFT of gravity, paying a price of arising an anisotropic scaling of time--space, $t\rightarrow\lambda^{z}t$, ${\bf x}\rightarrow\lambda{\bf x}$, thus Lorentz symmetry is broken by the presence of critical exponent $z$ but restored as $z$ flows to unity at low energies.

As in Ho\v rava--Lifshitz theories treating Lorentz invariance as an essentially low-energy phenomenon, in general we do adopt the point of view that Lorentz symmetry is still a good symmetry of any isolated QFT model on the flat space-time, and that its potential violations come from either placing the system in a ``medium'' or heat bath~(finite temperature), or coupling it to certain models of quantum gravity discussed above, which could also behave like a~(stochastic) medium, with nontrivial macroscopic consequences for the propagation of matter probes in such backgrounds. This results in modified dispersion relations and other nontrivial ``optical'' properties, as in the case of semiclassical backgrounds of loop gravity formalism~\cite{Gambini:1998it,Alfaro:1999wd,Alfaro:2002xz,Alfaro:2001rb,Li:2022szn} mentioned above.

Nonetheless, this sort of modification has been argued~\cite{Amelino-Camelia:2000stu,Amelino-Camelia:2002uql,Amelino-Camelia:2002cqb,Magueijo:2001cr,Magueijo:2002am,Zhang:2011ms} as well to characterize flat-space theories in some models in which the Planck ``length'' is considered a real length. As such it should be transformed under the usual Lorentz boosts. The requirement, that the Planck length becomes, along with the speed of light, \emph{observer independent}, leads quite naturally to modified ``Lorentz transformations'', and also to particle MDRs. This approach, dubbed ``doubly'' special relativity~(DSR)~\cite{Amelino-Camelia:2000stu,Amelino-Camelia:2002uql,Amelino-Camelia:2002cqb}, is a representative of the theories that have been denominated as new relativities with deformed kinematics; the others include the very special relativity of Cohen and Glashow~\cite{Cohen:2006ky}. In these cases Finsler geometry~\cite{Bao:2000rfg} often makes its appearance~\cite{Girelli:2006fw,Gibbons:2007iu,Zhu:2022blp,Zhu:2023mps}~(see~\cite{Zhu:2023kjx} for a recent review in this respect), and the same situation also characterizes, as we will discuss, the space-time~(D-particle) foam of string/brane theory~(c.f. Section~\ref{sec:4}). DSR should be distinguished from the other approaches mentioned so far, where Planck scale is associated to a ``coupling constant'' of the theory, being thus observer independent by definition. This is clearly the case of Einstein's gravity where $\Mp$ is related to the universal gravitational Newton's constant, and in~(critical and noncritical) superstrings, where the Planck length is related to the characteristic string scale of the theory, $\Ms$, which is also independent of inertial observers.

For our purposes in this review, we end at this point this brief but necessary digression of alternative theoretical models entailing non-Lorentz-invariant properties. Now let us get back to the case of strings.

\bi
\im Lorentz violation occurs spontaneously, thus is free from worldsheet conformal anomalies, in~(nonsupersymmetric, open) string field theory~\cite{Kostelecky:1988zi,Kostelecky:1989nt,Kostelecky:1991ak,Kostelecky:1995qk}, which implies nontrivial vacuum expectation values for certain tensorial quantities, as the perturbative string vacua are unstable. These are acceptable, in principle, string backgrounds, from a landscape viewpoint. A target-space EFT to consider phenomenology of such models is the renormalizable part~\cite{Colladay:1998fq,Colladay:1996iz} of the so-called standard-model extension~(SME)~\cite{Kostelecky:2003fs,Myers:2003fd,Mattingly:2008pw,Xiao:2008yu,Xiao:2010yx,Qin:2011md}. This latter framework is a systematic approach to incorporate general terms that violate Lorentz invariance at the action level by contracting operators of SM fields with controlling coefficients to preserve coordinate invariance~(diffeomorphism invariance, when gravity is incorporated~\cite{Kostelecky:2003fs}, in which case the coefficients would be dynamical). Albeit the phenomenology of possible LV from an EFT like this has been studied in the early days using the Coleman--Glashow proposal~\cite{Coleman:1998ti}, however this model is at most an illustrative scenario and cannot be taken as an ``exact'' theory, due to the requirement of physical-event consistency; see, e.g.,~\cite{Ma:2012dv}.\vspace{0.05cm}

Constructing nonrenormalizable~(called nonminimal SME) operators allows the inclusion of MDRs that entail Planck-mass suppression~\cite{Myers:2003fd,Mattingly:2008pw}. In cases where the theory is rotationally symmetric, as commonly assumed, the leading-order~(i.e., mass dimension-5) terms that consist of the modified quantum electrodynamics~(QED) with explicit Lorentz violation, also break CPT symmetry. In flat space it is implemented by introducing a Lorentz-noninvariant timelike 4-vector, $u^{\alpha}$ which describes the preferred frame, generated by QG, being associated with the vector~(tensor) condensate in the context of strings, for example:
\be\la{eq:d5a}
-\frac{\xi}{\Mp}u^{\mu}F_{\mu\alpha}u_{\nu}(u\cdot\pd)\tilde{F}^{\nu\alpha}+\frac{1}{2\Mp}\bar{\psi}\slashed{u}(\eta_{\textrm{L}}P_{\textrm{L}}+\eta_{\textrm{R}}P_{\textrm{R}})(u\cdot\pd)^{2}\psi.
\ee
Here $\xi$, $\eta_{\textrm{R,L}}$ are dimensionless SME coefficients, being phenomenological and of $\Or(1)$, $P_{\textrm{L,R}}=1/2(1\mp\gamma^{5})$ are left and right projection operators, and $\tilde{F}^{\mu\nu}=\frac{1}{2}\varepsilon^{\mu\nu\alpha\beta}F_{\alpha\beta}$ is the dual electromagnetic field strength. Equation~(\ref{eq:d5a}) yields contributions of $\Or(E/\Mp)$ to the dispersion relations of the photon~\cite{Myers:2003fd,Xiao:2009xe},
\be\la{eq:d5pdr}
\omega_{(\pm)}^{2}=\kv^{2}\pm\frac{\xi}{\Mp}2\omega^{2}\kn,
\ee
and for a fermion, such as the electron~(with the suffixes $\pm$ referring to helicity which can be shown to be a good quantum number in the presence of the LV terms~(\ref{eq:d5a})~\cite{Jacobson:2003bn}):
\be\la{eq:d5fdr}
E_{(\pm)}^{2}\simeq\pv^{2}+m_{e}^{2}+\eta_{\textrm{R,L}}\frac{E^{2}}{\Mp}\pn,
\ee
where $(E,\pv)$ is the 4-momentum of the particle, or $(\omega,\kv)$ for a photon~(with $\kv$ the wave vector and, $\omega$, its frequency), $m_{e}$ is the electron rest mass. For the antifermion, it can be shown by Dirac's ``hole interpretation'' arguments that the same dispersion holds, with $\eta_{\bar{q}_{\textrm{R,L}}}=-\eta_{q_{\textrm{L,R}}}$, where $\bar{q}$ and $q$ denote, respectively, antifermion and fermion~\cite{Jacobson:2005bg,Jacobson:2003bn}. We note that \emph{both} superluminal~($+$) and subluminal~($-$) photon propagation as well as the resulting characteristic birefringence effect are characterized by the same coefficient $\xi$. This will be crucially important when setting bounds to these CPT-odd LV couplings from photon decay~(see Section~\ref{sec:6}).
\im A rather different picture has emerged within~(the difficulties of) this rich formulation, in Liouville~(noncritical) string theory~\cite{Ellis:1992eh}, mentioned previously, whose development was partly motivated by intuition concerning the ``quantum gravity vacuum'' that is rather close to the one traditionally suggested by Wheeler~\cite{Wheeler:1964qna,Wheeler:1998vs} and subsequently adopted by Hawking~\cite{Hawking:1982dj}. They pioneered the idea that space-time at Planckian scales might acquire a foamy structure, which may thus behave like a dispersive ``\emph{medium}''. Evidence has been found in this approach~\cite{Amelino-Camelia:1996bln} as we now come to discuss, supporting the validity of MDRs~(refractive indices, etc), with the modification going linearly with the string length, $\Ls\equiv 1/\Ms$.\vspace{0.05cm}

This approach tackles background independence issues based on a~(perturbative) first-quantized framework, in such a way that it describes consistently background changes in string theory. As mentioned one should thus deal with $\sigma$-models away from their conformal~(fixed) points on the string-theory space. The noncriticality~(departure from equilibrium) of the string may be provided in concrete models for space-time foam, by a recoiling~(fluctuating) D-particle background~\cite{Ellis:1999uh,Ellis:1999jf,Ellis:2000sx,Ellis:1999rz}, playing the role of stochastic quantum-gravity foam. These examples will be explored in the next section while at present it is instructive to see how this may lead to MDRs which stem from \emph{spontaneous} Lorentz symmetry breaking within the generic Liouville strings~\cite{Ellis:1992eh} where the vertex operators deform the $\sigma$-model~(on the worldsheet $\Sigma$) as:
\be\la{eq:ncd}
\S=\S_{0}+g^{I}\int_{\Sigma}V_{g}(X).
\ee
Here, $X^{\mu}$ are target-space coordinates/$\sigma$-model fields, and $V_{g}$ are the vertex operators, corresponding to $g^{I}$. In stringy models of interest to us here, $g^{I}=\{G_{\mu\nu},\Phi,A_{\mu},\ldots\}$, i.e., it is a set of target-space background fields.\vspace{0.05cm}

The noncriticality of the deformation is expressed by nontriviality of the RG $\beta$-function of $g^{I}$, c.f.~(\ref{eq:bgf}): $\beta^{I}(g)=\sd g^{I}/\sd\ln\mu\neq 0$, where $\mu$, appearing above, is a local worldsheet scale, whose role, as mentioned before, may be played by the Liouville mode $\phi$ in light of the dynamical identification of the latter with the target time. Perturbation theory requires that one lies close to fixed point, which implies that one should work with $\beta^{I}$ expandable in power series in the couplings $g^{I}$. Quadratic order is sufficient for our purposes here, and to this order the $\beta$-function reads $\beta^{I}=(\D_{I}-2)g^{I}+c_{JK}^{I}g^{J}g^{K}+\ldots$, where no sum is implied in the first term, $\D_{I}-2$ is the anomalous scaling dimension, and $c_{JK}^{I}$ are operator product expansion~(OPE) coefficients. The theory is thus in need of gravitational~(Liouville) dressing~\cite{David:1988hj,Distler:1988jt,Distler:1989nt,Mavromatos:1989nf} in order to restore conformal symmetry and hence the consistency of the worldsheet theory~\cite{Green:2012oqa,Green:2012pqa}.\vspace{0.05cm}

For matter fields with central charge $\C_{m}>c^{\ast}$~(i.e., supercritical~\cite{Antoniadis:1988aa,Antoniadis:1988vi,Antoniadis:1990uu}), the Liouville field $\phi$ plays the role of an extra time coordinate in target space, as becomes evident from the dressed matter theory described by the following formulae:
\begin{align}\la{eq:ldma}
&\S_{\textrm{matter--Liouville}}=\S_{0}+\frac{1}{4\pi\Ls^{2}}\int_{\Sigma}\bigl[g^{I}(\phi)V_{g}-R^{(2)}\Q\phi-(\pd\phi)^{2}\bigr],\\
\la{eq:rcs}
&g^{I}(\phi)=g^{I}\e^{\alpha_{I}\phi}+\frac{\pi}{\Q\pm 2\alpha_{I}}c_{JK}^{I}g^{J}g^{K}\phi\e^{\alpha_{I}\phi},\\
\la{eq:domc}
&\alpha_{I}^{2}+\alpha_{I}\Q=-(\D_{I}-2),\quad\textrm{for supercritical strings},
\end{align}
where $g^{I}(\phi)$ are the dressed couplings, and $\alpha_{I}$ is the gravitational anomalous dimension. As mentioned above and demonstrated explicitly in certain toy models~\cite{Gravanis:2002gy}, one may identify the Liouville mode with time. Via noting the Liouville~(time) dependence of Equation~(\ref{eq:rcs}), one finds that one of the physical effects of such an identification is a modification of the time-dependence of the dressed $g^{I}$. For the case in the backgrounds of interest, i.e., of an almost flat space-time with small QG corrections, one may rewrite this relation as~\cite{Amelino-Camelia:1996bln}
\be\la{eq:rcsa}
g^{I}(\phi)\sim g^{I}(t)\e^{\I(\alpha_{I}+\D\alpha_{I})t},
\ee
where $\D\alpha_{I}$ depends on the OPE coefficients, $c_{JK}^{I}$, which encode the noncritical interactions of string probes with the ``environment'' of Planckian string states. For \emph{massless} low-energy string modes,
\be\la{eq:qgse}
\alpha_{I}\simeq\kn,\qquad\D\alpha_{I}\simeq\frac{\pi}{2\alpha_{I}}c_{JK}^{I}g^{K}=\Or\Bigl(\frac{E^{2}}{{\cal M}}\Bigr),
\ee
where the last equality originates in the OPE, which, for open strings, is of order $\Ls\kn$. This has been confirmed explicitly in the 2d string black-hole example~\cite{Ellis:1992eh,Ellis:1995nv}, and also characterizes D-particle ``foamy situations''~\cite{Ellis:1996dy,Kogan:1995df,Kogan:1996zv,Mavromatos:1998nz,Mavromatos:2001iz,Mavromatos:1998iu,Lizzi:1996fv,Ellis:1997cs} where, as we shall discuss below, there is \emph{minimal} suppression of the effects by a single power of the QG mass ${\cal M}$~(say, Planck mass $\Mp$ although this can be different from $\Ms$, which is a free parameter in string scenarios regarding our world as a membrane hypersurface~\cite{Polchinski:1998rr}~(``brane-world''; for a review, see~\cite{Maartens:2003tw})). From~(\ref{eq:rcsa}) one has wave dispersions and MDRs for massless string excitations:
\be\la{eq:glmdr}
\kn(\omega)\eqqcolon k\approx (1-\omega\Ls\varrho_{\textrm{L}})\omega,
\ee
which implies violation of Lorentz symmetry.\footnote{However, the reader should bear in mind that, in this approach this violation is spontaneous in the sense that the full string theory may be critical, and the noncriticality is only a result of restricting oneself in an effective low-energy theory~((foamy) ground state not respecting the symmetry).} The latter should not come as a surprise given the off-equilibrium nature of noncritical string which resembles an open system in which string matter propagates in a ``stochastic'' dissipative way, c.f.~(\ref{eq:gcyc})~\cite{Mavromatos:2002re}.\vspace{0.05cm}

The coefficient $\varrho_{\textrm{L}}$ and its sign are to be determined in specific models of quantum foam elaborated in the next section, involving stringy interactions of low-energy matter with the recoiling D-particle defects as the main cause of departure from criticality~(conformality of the associated $\sigma$-model background). These models have interesting features, including the \emph{subluminal} propagation of photons in the corresponding QG medium, that could accommodate~\cite{Li:2021gah} most of the exclusion limits, as well as the potential hint, of LV from astrophysical data~\cite{Shao:2009bv,Zhang:2014wpb,Xu:2016zxi,Xu:2016zsa,Amelino-Camelia:2017zva,Xu:2018ien,Liu:2018qrg,Chen:2019avc,Zhu:2021pml,Zhu:2021wtw,Zhu:2022usw,He:2022gyk,Song:2024and,Song:2025qej,Song:2025apj,Amelino-Camelia:2016ohi,Huang:2018ham,Huang:2019etr,Huang:2022xto,Song:2025akr}~(c.f. Section~\ref{sec:5}).
\ei

\section{Lorentz and CPT in String/Brane-World Inspired D(efect)-Foam}\label{sec:4}

We discuss below a model for stringy space-time foam~(or better, a class of such models) initially inspired from the above-exposited Liouville formalism in the modern framework of the D-brane approach to QG~\cite{Polchinski:1995mt,Polchinski:1996na}. Such models have been dubbed \emph{stringy/D-brany defect foam}~\cite{Ellis:1999sf,Ellis:2000sf,Ellis:2004ay,Ellis:2005ib,Mavromatos:2007xe,Ellis:2008gg,Ellis:2009vq,Mavromatos:2010pk,Li:2021gah,Li:2021eza,Li:2022sgs,Li:2023wlo,Li:2024crc,Li:2009tt} or D-foam for short, consisting of a bulk ``punctured'' with localized space-time D-particle defects. The latter are either allowed background configurations as in~(supersymmetric) type IIA or IA~\cite{Schwarz:1999xj}~(a T dual of type I) string theory or arise effectively from suitable compactifications of higher-dimensional branes~(e.g., 3-branes wrapped up in appropriate compact spaces in the context of IIB strings~\cite{Li:2009tt}). The presence of these defects which string matter probes might be scattered by, implies nontrivial optical properties for the QG vacuum, in the sense of particle-energy dependent indices of refraction. Such effects may be relevant for LV searches from high-energy astrophysics that shall be discussed later on. Below we review briefly the main results.

The cosmological framework of the approach is based on appropriate stacks of parallel brane-worlds~\cite{Maartens:2003tw}, some of which are moving in a higher-dimensional bulk space where a finite density gas~(medium) of defects is distributed. Supersymmetry dictates the number of parallel D-branes allowed to appear in each stack but does not restrict the concentration of defects in the bulk. In the limit of static branes and D-particles this configuration constitutes an appropriate supersymmetric ground state of this brane theory~\cite{Ellis:2004ay}, while bulk motions of the branes result in nonzero ``vacuum''~(or, rather, ``excitation'') energy, and hence the breaking of target-space supersymmetry. Our observable Universe is represented allegedly by one of these roaming branes, compactified to three spatial dimensions~(D3-brane). As it sweeps through the bulk, the \emph{randomly} traversing D-particle defects appear as an effective ``quantum foam'', c.f. D-foam, from the brane observer viewpoint. Below, we consider it to be present in string theories of phenomenological interest as in type I/IIA framework~\cite{Ellis:2004ay,Ellis:2005ib,Ellis:2008gg}, since, even if it is not allowed to exist, there can be effective D-``particles'' as mentioned above, constructed out of wrapped-up type-IIB 3-branes~\cite{Li:2009tt}.

Within such microscopic I/IIA D-particle models the defects break Poincar\'e invariance, and their recoil following scattering off the matter strings violates Lorentz invariance locally. As a result of charge conservation, charged~(open) stringy matter excitations cannot interact with these D0-brane defects, which are truly pointlike and electrically neutral. Only \emph{charge neutral} particles are interacting predominantly~(only, gravitationally) with the foam~(due to the weakly interacting character of the defect). Cosmologically, therefore, it is photons and neutrinos that ``feel'' mostly the effects of special QG configurations that modify their in-vacuo propagation. It is this nonuniversality that implies the \emph{violation of the equivalence principle}, in the sense that different relativistic particle species break Lorentz symmetry with varying amounts~\cite{Ellis:2003ua,Ellis:2003sd,Ellis:2003if}. Especially, charged leptons such as electrons would propagate in a Lorentz-invariant way, \emph{in contrast} to photons~\cite{Li:2025ste}.

Topologically nontrivial interactions between these defects with open strings, anchored on the brane Universe and representing neutral matter, involve temporary capture/splitting and subsequent reemission of the particles by the defects. Intermediate strings are created during capture~\cite{Ellis:2004ay,Mavromatos:2010nk} which is represented from a worldsheet point of view as an \emph{impulse}. Local distortions of the neighboring space-time due to the recoil of massive defects can be calculated perturbatively, for weakly coupled strings, using logarithmic CFTs~\cite{{Mavromatos:1998nz,Mavromatos:1998iu},Mavromatos:2004bx}. The relevant $g^{I}=\V_{i}$~($i=1,2,3$, as we are interested in recoil components \emph{along} our 3-brane) so that the pertinent vertex operator reads
\be\la{eq:rvo}
V_{\textrm{imp}}=\int_{\pd\Sigma}\varepsilon(\varepsilon Y_{i}+\U_{i}X^{0})\Theta_{\varepsilon}(X^{0})\pd_{n}X^{i},
\ee
where $Y_{i}$ is a collective spatial coordinate of the D-particle and $\Theta_{\varepsilon}$ is a regularized operator:
\be\la{eq:rhf}
\Theta_{\varepsilon}(t=X^{0})=-\frac{\I}{2\pi}\int_{-\infty}^{\infty}\frac{\sd q}{q-\I\epsilon}\e^{\I qt},\quad\textrm{with}~\varepsilon\rightarrow 0^{+}.
\ee
Above $X^{0}$ is the target-space time coordinate, obeying Neumann boundary conditions while $X^{i}$, the spatial coordinates on the brane, satisfy Dirichlet boundary conditions. The notation $\pd_{n}$ denotes normal derivative on the worldsheet. By $\U_{i}$, we denote the recoil relativistic 3-velocity of the brane-confining D-particle of mass $\Ms/\gs\eqqcolon\Md$~(where $g_{s}<1$ is the~(weak) string coupling): $\U_{i}=\gamma_{\V}\V_{i}$, with $\gamma_{\V}=(1-\bfV^{2})^{-1/2}$, the corresponding Lorentz factor, but for a heavy, nonrelativistic D0-brane where $\V_{i}$ is small it is well approximated by the ordinary velocity as given by
\be\la{eq:drv}
\V_{i}\frac{\Ms}{\gs}=\D k_{i}\equiv \zeta^{(i)}k_{i}.
\ee
The~(dimensionless) variables $\zeta$ appearing above, are related to the fraction of momentum $\D k_{i}$ that is transferred at a collision with a D-particle, provided the recoil process conserves momentum, as demonstrated rigorously in~\cite{Kogan:1996zv,Mavromatos:1998nz,Mavromatos:2001iz}, following the closure of a worldsheet logarithmic algebra.

The impulse operator in Equation~(\ref{eq:rvo}) has anomalous scaling dimension $-\varepsilon^{2}$/2. Its presence drives the stringy $\sigma$-model \emph{supercritical} and thus Liouville dressing is required~\cite{David:1988hj,Distler:1988jt,Distler:1989nt,Mavromatos:1989nf}. This latter procedure results in the recoil velocity part of the deformation~(\ref{eq:rvo}) being replaced by the worldsheet bulk operator~\cite{Ellis:1999uh,Ellis:1999jf,Ellis:2000sx}:
\be\la{eq:brbo}
V_{\textrm{bulk}}\supset -\varepsilon^{2}\V_{i}\int_{\Sigma}\e^{\alpha_{L}\phi}t\Theta_{\varepsilon}(t)\pd^{\beta}\phi\pd_{\beta}X^{i},
\ee
where $\beta=1,2$ is a worldsheet index, $\alpha_{L}$ is the Liouville anomalous dimension, which is, to leading order in $\varepsilon$~(c.f.~(\ref{eq:domc})), $\alpha_{L}\sim\varepsilon$, whilst the central charge \emph{surplus}, $\Q^{2}>0$, is of order $\varepsilon^{4}$. Upon identifying the zero modes of the Liouville with the target time in~(\ref{eq:brbo})~\cite{Ellis:1997cs}, using the representation $\Theta_{\varepsilon}(X^{0})\sim\vartheta(t)\e^{-\varepsilon t}$, where $\vartheta(t)$ is the ordinary Heaviside step function, and considering relatively large times $t\gg 0$ after the scattering event, $\varepsilon^{2}t\vartheta(t)\sim 1$, one obtains an off-diagonal deformation $h_{\mu\nu}$ of the target-space geometry, $g_{\mu\nu}=\eta_{\mu\nu}+h_{\mu\nu}$~\cite{Ellis:1999jf,Ellis:1996fi} with $0i$ components,
\be\la{eq:flm1}
g_{0i}=h_{0i}(X^{0},\V_{i})\sim\varepsilon^{2}\V_{i}t\vartheta(t)\e^{-2\varepsilon t}.
\ee
For times of order of the duration, $\varDelta t$, of the impact~(collision), $X^{0}\sim\varDelta t\sim 1/\varepsilon$, the order of magnitude of $g_{0i}$ is~\cite{Ellis:1999sd},
\be\la{eq:flm2}
g_{0i}(X^{0},\V_{i})\sim\V_{i}.
\ee
This expresses the back-reaction effects of the recoiling heavy fluctuating D-particle defect(s) on the surrounding space-time. Note, that the distortion of target space is by definition \emph{local}, and applies only in the neighborhood of the defect, for the perturbative $\sigma$-model formalism used here to derive such an effect~(\ref{eq:flm2}).

The outgoing scattered matter string, then ``sees'' the distorted space-time~(\ref{eq:flm2}) in which the Lorentz breaking effects are proportional to $\V_{i}$, as the D-particle recoils when struck by the incident particle. The important feature that, the recoil-induced geometry perturbations depend on the energy content, or momentum of the probe~(Finsler type~\cite{Bao:2000rfg}) does not arise in other LV approaches to quantum space-time, and will be important in the formulation of microscopic models of D-foam. In general, the associated effects depend on the density of D-particles encountered by the string state and, we represent such foam-averaged effects via $\sdla\cdots\sdra_{\Dp}$, which denotes \emph{both} statistical averages over populations of defects~\cite{Ellis:2004ay}, as well as target-space quantum fluctuations~\cite{Ellis:1999jf,Ellis:2000sx} coming from the resummation over worldsheet genera~\cite{Mavromatos:1998nz,Mavromatos:1998iu}. These, at present, are not treated systematically and so we revisit here, for concreteness and relevance to our discussion below, three different proposals, in particular their predictions for the nonstandard optical properties for the brane, i.e., refractive indices of~(observable) matter localized on the brane-world.

\bi
\im[(1)] \underline{Gravity-Induced Dispersion in the Anisotropically-Recoiling Foam}\vspace{0.15cm}

In an early development of the approach~\cite{Ellis:2004ay} metric perturbations caused by different D-particles ``sum'' up all together such that, from the~(macroscopic) point of view of an observer confined to our brane-world~(as real observers are regarded in this picture), a back-reacted effect due to the foam, viewed as a whole, occurs on average, $\sdla\V_{i}\sdra_{\Dp}\neq 0$, implying that there is \emph{anisotropy} in the direction of the average recoil velocity of the defect, oriented along that of the propagating incident string. For gauge degrees of freedom, say, photons, the dynamics of recoil excitations has a Born--Infeld~(BI) form of nonlinear ``electrodynamics''~\cite{Kogan:1995df,Kogan:1996zv,Mavromatos:1998nz,Mavromatos:2001iz,Mavromatos:1998iu,Lizzi:1996fv} requiring an average \emph{deceleration} of the particle by the foam, $\sdla\kv\cdot\bfV\sdra_{\Dp}=-\sdla k\lvert\bfV\rvert\sdra_{\Dp}<0$. Subluminal~(group) velocities, suppressed minimally by a single power of $\Ms/\gs$~\cite{Ellis:2004ay,Li:2021gah,Li:2021eza},
\be\la{eq:afpv}
c_{\textrm{gr}}\simeq 1-2\gs\frac{\zeta_{\Dp}\kn}{\Ms}\sim 1-\lvert\Or(n_{\Dp}\omega/\Md)\rvert,\quad\sdla\zeta\sdra_{\Dp}\eqqcolon\zeta_{\Dp}>0,
\ee
obtained after averaging over foam populations, arise from $k_{\mu}k_{\nu}g^{\mu\nu}=-E^{2}+2k^{i}\V_{i}E+k^{i}k_{i}=0$, for which the positive energy solution $k^{0}=\omega>0$, connecting smoothly with the case of foam-free vacuum~(no recoil $\V_{i}\rightarrow 0$), reads
\be\la{eq:gfmdr}
\omega=\kn (1+(\bfV\cdot\hat{\kv})^{2})^{1/2}+\kv\cdot\bfV,
\ee
where $\hat{\kv}=\kv/\kn$~(prior to averaging over populations of D-particles). We assumed that the relation $v=\pd E/\pd\pn$ holds although this is in dispute and at least one concrete stringy example has been provided that it fails~(see below), and, for simplicity, we also assumed \emph{isotropic} foam that requires $\zeta^{(i)}=\zeta$ for all $i=1,2,3$, in~(\ref{eq:drv}). In Equation~(\ref{eq:afpv}) we omit $\sdla\cdots\sdra_{\Dp}$ in the notation for $k_{i}$ and $\omega$~(since they all denote such averages); $n_{\Dp}$ is the effective~(linear) density of defects on our brane, a microscopic free parameter in the theory besides the D-particle mass. The effective mass scale that suppresses the corresponding effect of LV refraction, $\eta-1=\Or(E/\Ms)$, which notably is of the form implied by~(\ref{eq:glmdr}) with $\varrho_{\textrm{L}}<0$, is~\cite{Li:2021eza}
\be\la{eq:afqg}
M_{\textrm{sQG}\gamma}\coloneqq\frac{\Md\gs}{\varsigma_{\Dp}},\quad\textrm{where}~\Or(1)\gtrsim\varsigma_{\Dp}\coloneqq 2\gs\zeta_{\Dp}.
\ee
It is thus arbitrarily free. Testing whether higher-energy photons are \emph{retarded} relative to lower-energy ones, rather than advanced, with astrophysical $\gamma$-rays from, say, GRBs, provides important constraints on the scale.\vspace{0.05cm}

The D-particle Born--Infeld analysis of recoil has been extended to the case of photinos in~\cite{Ellis:1999sf} via an appropriate supersymmetrization of the treatment. Further, we recently observed~\cite{Li:2024crc} that upon averaging the result thereof~(for a single scattering event) over this \emph{anisotropic-recoil} foam, involving a collection of D-particles,\footnote{Their concentration is constant in time in models where \emph{uniform} D-particle backgrounds are considered, as we do here, for simplicity, although it could vary with the cosmological epoch~\cite{Ellis:2009vq}. The fact that the effective QG scale(s) would depend on the redshift $z$, $M_{\textrm{sQG}}\rightarrow M_{\textrm{sQG}}(z)\sim\frac{\Ms}{\gs n_{\Dp}(z)}$, further complicates the situation.
} we expect subluminal propagation also for the~(almost massless) neutrinos with the refractive indices being akin to that of the bosonic matter just discussed, without any chirality dependence:
\be\la{eq:afnri}
\eta\simeq 1+\frac{\kn}{M_{\textrm{sQG}\nu}}>1,\quad\textrm{for both}~\nu\textrm{'s and}~\bar{\nu}\textrm{'s},
\ee
where the gravity foam mass of neutrinos, $M_{\textrm{sQG}\nu}$, \emph{does not necessarily} match, however, that of photons in magnitude, despite their natural order $\Or(\Mp)$. Indeed, we mention that such foam effects need not obey the principle of equivalence in the sense of being universal to all particle types~\cite{Ellis:2003ua,Ellis:2003if}.\vspace{0.05cm}

Thus far we have learnt that strings exhibit background independence, within a perturbative~(noncritical) $\sigma$-model formulation~\cite{Kogan:1996zv,Ellis:1996fi}, where via~(\ref{eq:drv}), the summation over worldsheet genera elevates the recoil-$\V_{i}$ field into a quantum(-mechanical) operator: $\sdla\V_{i}\sdra_{\Dp}\Rightarrow -\I\zeta_{\Dp}\nabla_{i}/\Md$~\cite{Mavromatos:1998nz,Ellis:1995nv}. Via this prescription for first quantization, it can be shown that the BI effective action is mapped to a local effective Lagrangian responsible for the modified Dirac~(Maxwell's) equation for the neutrino~(radiation) fields, and the respective MDRs~(c.f.~(\ref{eq:afnri}) and~(\ref{eq:afpv}), above). This accordingly requires the momentum of the particle met by a D-particle $k\ll\Ms$, assumed implicitly as we expanded~(\ref{eq:gfmdr}) in powers of $\V_{i}$ which is small.
\im[(2)] \underline{Uncertainty-Caused Delay from Stretched-String Formalism}\vspace{0.15cm}

Since the scale $\Ms$ is arbitrary from a modern perspective and for instance, in low-scale string/brane~(cosmology) models~(with large extra-dimensions)~\cite{Antoniadis:1999rm,Antoniadis:2000vd}, it can be of $\Or(10~\textrm{TeV})$~(though, it cannot be lower than this because if this were the case we should have already seen fundamental strings experimentally), it is possible that $E\sim\Md$ and the effective low-energy field theory formulation would no longer be valid for energies above this value~\cite{Burgess:1986dw}. In extending this prescription beyond such a classical EFT so as to describe some aspects of the foam for arbitrary momenta, in~\cite{Ellis:2008gg} a nonperturbative mechanism for the vacuum refraction-induced photon delays was proposed, based on the pertinent \emph{stringy uncertainty principles}, particularly the time--space uncertainties of the string~\cite{Yoneya:2000bt},
\be\la{eq:sucp}
\D t\D X\geq\alp.
\ee
where $\alp\equiv\Ls^{2}$ is the universal Regge slope. The microphysical modeling of the matter--defect interaction in~\cite{Ellis:2008gg,Ellis:2009vq} shows that the intermediate string~(and hence nonlocal) state, formed in this process and \emph{stretched} between the D-brane-world and the defect, exhibits length oscillations as it stores the incident energy $k^{0}$ of the particle as potential energy. The time spent~(delay) for this string to grow to the maximal length and then shrink back to zero size is
\be\la{eq:dtdn}
\D t\sim k^{0}/\Ms^{2}>0,
\ee
which as argued in~\cite{Ellis:2008gg,Mavromatos:2010pk}, is a direct consequence of~(\ref{eq:sucp}), describing the delay caused in each collision of a photon~(or neutrino) state with a D(-particle)-defect. The situation may be thought of as the stringy/brany analogue of wave propagation in conventional media, where the role of the electrons~(modeled as harmonic oscillators) is played here by space-time D-defects.\footnote{Contrary to that case however, the string-inspired refraction effect~(c.f.~(\ref{eq:sfri}) below) is proportional to the energy, while the effective scale that suppresses the effect~(\ref{eq:dtdn}) is the QG~(string) scale and not the electron mass in the familiar local field theory description of the phenomenon in solids.} The inclusion of recoil of the latter results in higher-order corrections of the form $\D t\sim E\alp/(1-\lvert\bfV\rvert^{2})$, obtained by analogy of the case with that of open strings in an external electric field~\cite{Seiberg:1999vs,Seiberg:2000ms,Seiberg:2000gc}. Despite the \emph{noncommutativity of space-time} from this analogy, $[X^{i},t]\sim\V^{i}$, the uncertainty-related delays~(\ref{eq:dtdn}) are purely stringy effects and cannot be captured by noncommutative local effective field theories which suffer from noncausal effects. Such delays, being independent of helicities~(i.e., \emph{no} birefringence in photon propagation), respect causality.\vspace{0.05cm}

It was shown, next, in~\cite{Ellis:2008gg,Ellis:2010he,Li:2024crc} that the overall retardation experienced by a particle interacting nontrivially with a foam that has a linear density of defects $n_{\Dp}/\sqrt{\alp}$, via the above-mentioned splitting, is proportional to $En_{\Dp}\Ls$. This corresponds to an effectively LV index of refraction \emph{in vacuo}:
\be\label{eq:sfri}
0<\eta(E)-1\simeq n_{\Dp}\alp E\sigma_{\Dp}\Bigl[\frac{\gs}{\Ls}\Bigr]\ll 1,
\ee
where $\sigma_{\Dp}$ is the matter--D-defect scattering cross section. As argued in~\cite{Li:2024crc}, such effects characterize not only photons but neutrinos~(antineutrinos)~(or, Majorana neutrinos of both chiralities) in this foam. The new feature here is that~\cite{Ellis:2008gg} the capture/reemission process is described within the language of critical string theory. Novelties which have important consequences for phenomenology, as we shall mention later on, in this case include~\cite{Ellis:2010he,Mavromatos:2010pk}:
\bi
\im[$\divideontimes$] The \emph{disentanglement} of the linear-in-energy time lag~(\ref{eq:dtdn}) from MDRs, given that the model of~\cite{Ellis:2008gg} does not entail any modification of the local dispersion relations of particles~(for which, the averaged Finsler dispersion corrections are proportional to $\V_{i}\V^{i}$~\cite{Li:2024crc},
\be\la{eq:qsdr}
E^{2}-\kv^{2}\sim\Or\Bigl(+\,\kv^{2}\gs^{2}\frac{E^{2}}{\Ms^{2}}\Bigr).
\ee
with the r.d.s. being \emph{quadratically} suppressed by the string~(mass) scale if recoil of the defect is \emph{isotropic}, $\sdla\V_{i}\sdra_{\Dp}=0$; this case may thus be thought of as implying an example of a possible failure of the velocity-energy relation, mentioned previously, in the contexts of QG), and:
\im[$\divideontimes$] The \emph{destabilization of the vacuum}~\cite{Burgess:1986dw}, when the D-defect recoil velocity approaches the relativistic velocity of light, $\lvert\bfV\rvert^{2}=\Or(1)$, as can be attained in low-scale string models~\cite{Antoniadis:1999rm}, in which it is possible that the particle energy $k\sim\Ms/\gs$. In such cases, $\D t\rightarrow +\infty$ implies that particles of such energies would be \emph{destabilized}~(absorbed) when scattering off the D-particle, which strongly curves the surrounding space-time and behaves as a result like a black hole, capturing permanently the incident particle.
\ei
\im[(3)] \underline{A Stochastic, Isotropic Foam with CPT-Violating Neutrinos}\vspace{0.15cm}

At this point we remark that, in \emph{isotropic-recoil} string foam situations where $\sdla\V_{i}\sdra_{\Dp}=0$, Lorentz invariance may be conserved on average, and is lost only via considering recoil velocity fluctuations among statistical ensembles of defects, namely, foam types~\cite{Li:2022sgs,Li:2023wlo} where one assumes Gaussian stochastically fluctuating configurations taking $\zeta$'s in~(\ref{eq:drv}) as \emph{random} variables, via the moments~\cite{Mavromatos:2005bu}:
\be\label{eq:sfc}
\sdla\zeta^{(i)}\sdra_{\Dp}=0,\qquad\sdla\zeta^{(i)}\zeta^{(j)}\sdra_{\Dp}=\bigl(\fd_{\Dp}^{(i)}\bigr)^{2}\delta_{ij}\neq 0,
\ee
i.e., even powers of $\V_{i}$ yield nonzero population averages. In an isotropic foam case, as we assumed here, $\fd_{\Dp}^{2}\coloneqq\sdla (\zeta^{(i)})^{2}\sdra_{\Dp}\vert_{i=1,2,3}$ along all spatial directions. The variance, $\fd_{\Dp}^{2}$, is free, albeit small~(and, naturally up to $\Or(1)$), to be determined phenomenologically, as it is a function of the density of the foam, $n_{\Dp}$.\vspace{0.05cm}

The key features of the proposal in~\cite{Li:2022sgs,Li:2023wlo}, that discriminate it from the above scenarios, include that the neutrino in-vacuo propagation, due to the stochastic effects of the foam, becomes entangled with the particle/antiparticle nature of neutrinos. We mention that the \emph{restoration} of Lorentz invariance, due to the foam-background stochasticity~(\ref{eq:sfc}), is a desirable feature~\cite{Mavromatos:2010nk} for a~(string) model. In this case albeit the symmetry violations, expressed via the back-reacted metric~(\ref{eq:flm2}), are washed out \emph{statistically} and isotropy and rotational invariance are preserved, nevertheless the kinematical aspects of scatterings of~(anti)neutrinos, of mass $m_{\nu}$, with D-defects are at play, via~\cite{Mavromatos:2012ii},
\begin{align}\la{eq:sec}
E&=E^{\prime}+\Bdla\frac{1}{2}\Md\bfV^{2}\Bdra_{\Dp}+\ldots,\\
\la{eq:smc}
\kv&=\kv^{\prime}+\Md\sdla\bfV\sdra_{\Dp}=\kv^{\prime},
\end{align}
where, $\frac{\Ms}{2\gs}\bfV^{2}$ is the kinetic energy of a defect, and $(E^{\prime},\kv^{\prime})$ is the averaged 4-momentum of the reemitted particle. In~(\ref{eq:sec}), $\ldots$ denote corrections that are negligible in the case of nonrelativistic recoil, and $E$, obtained from $k_{\mu}k_{\nu}g^{\mu\nu}=-m_{\nu}^{2}$~(from which however the negative $E$ solution should be kept now), stands for the average energy for neutrino species. Then, the outgoing~(average) energy, $E^{\prime}$, yields asymmetrically MDRs between neutrinos and antineutrinos~\cite{Li:2022sgs,Li:2023wlo}:
\be\la{eq:sfndr}
E\simeq\lvert E_{\textrm{M}}\rvert-\fd_{\Dp}^{2}\frac{\kv^{2}}{2\Md},\quad\textrm{for}~\nu\textrm{'s};\quad E\simeq\lvert E_{\textrm{M}}\rvert+\fd_{\Dp}^{2}\frac{\kv^{2}}{2\Md},\quad\textrm{for}~\bar{\nu}\textrm{'s},
\ee
where we define $E_{\textrm{M}}=\pm\sqrt{\kv^{2}+m_{\nu}^{2}}$ as the Minkowski energy of an \emph{indefinite} signature. Superluminality of antineutrinos is then established, as evidenced from their velocities, i.e., $\delta=\pd E/\pd\kn-1\simeq\kn/\mathring{M}_{\textrm{sQG}\nu}$, where the suppression scale $\mathring{M}_{\textrm{sQG}\nu}\coloneqq\Ms/(\fd_{\Dp}^{2}\gs)$, whilst for neutrinos $\delta\simeq -\fd_{\Dp}^{2}\kn/\Md<0$ and hence the neutrino propagation is CPT-violating.\vspace{0.05cm}

We remark at this stage that, the in-vacuo dispersion for $\bar{\nu}$'s in~(\ref{eq:sfndr}) is obtained, $E(\kv)\coloneqq -E^{\prime}>0$, on the assumption that the antifermions have negative energies following the ``hole theory''
of Dirac. These considerations can only apply to fermions, as opposed to bosons. The speed of energetic photons is slower than lower-energy ones, due to their bosonic property in this scenario. The corresponding MDR is thus identical in form to the first expression of Equation~(\ref{eq:sfndr})~\cite{Li:2025ste}.\vspace{0.05cm}

For our purposes in this work it is also important to remark that for the explicit models discussed~\cite{Li:2022sgs,Li:2023wlo} the formalism of local effective Lagrangians breaks down in the sense that~\cite{Ellis:2000sf}, during \emph{interactions}, total energies of the observable particles \emph{are not in general} conserved, in contrast to the case of propagation. For this reason, caution should be exercised in testing such models using the result derived from most phenomenological studies, where energy conservation is usually assumed implicitly. It is based upon this assumption that several anomalous processes, such as $\nu\rightarrow\nu ee^{-}$ or $\nu\rightarrow\nu\nu\bar{\nu}$, which would render the superluminal~(anti)neutrinos unstable~\cite{Cohen:2011hx}~(c.f. some discussion in Section~\ref{sec:6}), were used to limit neutrino superluminality. However, in such 4-particle interactions, which may be factorized as products of 3-particle vertices mediated by the exchange of an intermediate particle excitation with a momentum $\pv$, the energy is \emph{lost} in the presence of the fluctuating~(in target-space) D-particle foam by an amount of $\sim (\varsigma_{I}/\Md)\pv^{2}$, where $0<\varsigma_{I}$, which $\neq\fd_{\Dp}^{2}$ in general, controls the intensity of the loss in the reaction. The analysis in~\cite{Li:2022sgs,Li:2023wlo} shows that the presence of such losses modifies the relevant energy thresholds $E_{\textrm{th}}$, as,
\begin{align}\la{eq:fpcth}
E_{\textrm{th}}&\simeq\Bigl[4m_{e}^{2}\frac{\Ms}{\gs(\fd_{\Dp}^{2}-4\varsigma_{I})}\Bigr]^{1/3},\quad\textrm{for}~\bar{\nu}\rightarrow\bar{\nu}ee^{-},\\
\la{eq:fnsth}
E_{\textrm{th}}&\simeq\Bigl[9m_{\nu}^{2}\frac{\Ms}{\gs(\fd_{\Dp}^{2}-2\varsigma_{I})}\Bigr]^{1/3},\quad\textrm{for}~\bar{\nu}\rightarrow\bar{\nu}\nu\bar{\nu},
\end{align}
for the reactions to occur, which stem from kinematics, in such a way that the otherwise stringent bounds imposed from observations of ultra-high energy~(UHE) astrophysical neutrinos ($E\gtrsim 100$~TeV), are evaded. Indeed, as we can see, for example, from~(\ref{eq:fpcth}), the process is forbidden in vacuum if $\fd_{\Dp}^{2}\leq 4\varsigma_{I}$, or, effectively inhibited if $\varsigma_{I}\approx (\fd_{\Dp}/2)^{2}$ in case $\fd_{\Dp}^{2}>4\varsigma_{I}$, hence in the present model, generically, superluminal antineutrinos do not suffer from the problem of~(in)stability.
\ei
We conclude the theoretical part of the present review by recalling that the light propagating in space-time D-foam entails a linearly frequency/energy-dependent refractive index, that is \emph{always} subluminal, resulting in a negative variation in the light speed with particle energy, and hence, a delay in the propagation of higher-energy photon over a distance $L$ \emph{in vacuo} as compared to softer ones,
\be\la{eq:dft}
\Delta t^{\textrm{D-foam}}\sim L\delta(k^{0})\sim\frac{E}{\Ms}n_{\Dp}L,
\ee
although the origin of such lags differs among D-particle models. The possibility of birefringence, namely, a difference between the velocities of light with different polarizations~(see below for details) does not arise in this approach. The effect for neutrino propagation could be foam-model dependent, since energetic neutrinos~(antineutrinos) may propagate \emph{either}, subluminally~\cite{Ellis:1999sf,Li:2024crc,Ellis:2008gg}, or in a way breaking CPT. In this latter case~\cite{Li:2022sgs,Li:2023wlo}, superluminal antineutrinos are ``protected'' by the induced energy nonconservation as they interact in the stochastic D-foam quantum-fluctuating backgrounds~\cite{Mavromatos:2005bu}. D-foam is also \emph{flavor blind}~\cite{Ellis:2001nxa}, evading tight limits from oscillation physics~\cite{IceCube:2017qyp} that would be applied in LV models with flavor~(i.e., neutrino-species) dependent modifications to neutrino dispersions. As it will be clear in the next sections, these features render the whole approach quite attractive, when confronted with various LV tests in high-energy astrophysics.

\section{Searches for Energy-Dependent Time-of-Flight Lags}\label{sec:5}

In attempt to encapsulate QG phenomenology, energy-dependent propagation velocity of relativistic particles, of the form introduced at the beginning of our work~(c.f.~(\ref{eq:mdr})), has a broad scope to bound LV, and in general violation of fundamental symmetries, due to exotic effect of the microscopic fabric of space-time. As we can see, despite being quite different, a plethora of Lorentz- and/or CPT-violating frameworks for QG agree in predicting such a speed variation~\cite{Jacobson:2004rj,Mattingly:2005re,Jacobson:2005bg,Liberati:2009pf,Shao:2009bv,Zhang:2014wpb,Liberati:2013xla,Ellis:2005sjy,Xiao:2009xe}. This implies, in general contexts of LV, modifications on the in-vacuo dispersions for the particles; however, again, we emphasize that it may not be so concerning particular QG proposal, as is for instance the case of the stretched-string D-foam~\cite{Ellis:2008gg,Ellis:2009vq,Mavromatos:2010pk,Li:2024crc}. In such a case the \emph{causal} delays of the photons~(neutrinos) emerging after capture by the defect, scaling \emph{linearly} with the particle energy, are purely stringy effects, associated with time--space uncertainties and not directly with MDRs, whose corrections, for the most physically intriguing case of no average recoil, $\sdla\V_{i}\sdra_{\Dp}=0$, are quadratically suppressed by the string scale, c.f.~(\ref{eq:qsdr}).

Nonetheless, in the presence of LV effects, photons and/or high-energy neutrinos~(with negligible masses $m_{\nu}\lesssim 1$~eV~\cite{Huang:2014qwa}) with different energies should spend slightly different time-of-flight to reach the observer, regardless if or not the variation~(\ref{eq:mdr}) can be interpreted as a traceable result of MDRs. This plays an indispensable role in the search of LV signatures in astroparticles because of the high energies and the gigantic propagation distances. The technique, known as time-of-flight~(TOF) test, as was first suggested in~\cite{Amelino-Camelia:1996bln,Amelino-Camelia:1997ieq}~(and in~\cite{Jacob:2006gn}), has been extensively studied in literature~(\cite{Ellis:1999sd,Ellis:2002in,Ellis:2005sjy,Ellis:2008fc,Amelino-Camelia:2009imt,Xiao:2009xe,Wang:2016lne,Ellis:2018ogq,Ellis:2018lca}, to mention a few). Various observations of the sharp energetic signals from distant sources, such as GRBs~(as registered by the Fermi satellite~\cite{Fermi-LAT:2009owx,FermiGBMLAT:2009nfe,Ellis:2018lca}, and other ground-based telescopes~\cite{MAGIC:2020egb,MAGIC:2019lau}), pulsars~\cite{MAGIC:2017vah} and active galactic nuclei~(AGNs), or blazars~\cite{Biller:1998hg,MAGIC:2007etg,HESS:2008irp,HESS:2011aa,HESS:2019rhe,Li:2020uef}, have already achieved the required sensitivity to the propagation effect induced by QG.

Over the past decade or so, dedicated exploration for a possible signature of anomalous light-speed variation has been carried out using the time lags in the arrival of cosmic $\gamma$-rays coming from GRBs~\cite{Shao:2009bv,Zhang:2014wpb,Xu:2016zxi,Xu:2016zsa,Amelino-Camelia:2016ohi,Amelino-Camelia:2017zva,Xu:2018ien,Liu:2018qrg,Chen:2019avc,Zhu:2021pml,Zhu:2021wtw,Zhu:2022usw,He:2022gyk,Song:2024and,Song:2025qej,Song:2025apj,Song:2025akr} and AGNs~\cite{Li:2020uef}. It is such kind of studies as well as an analogous statistical signal~\cite{Amelino-Camelia:2016fuh,Amelino-Camelia:2016wpo,Amelino-Camelia:2016ohi,Amelino-Camelia:2017zva,Huang:2018ham,Zhang:2018otj,Huang:2019etr,Huang:2022xto} in the propagation of cosmic neutrinos that we shall restrict our attention to for the purposes of this review. We shall try to see what these results imply for the QG models of stringy~(D-particle) space-time foam discussed above, and how the latter compare with other LV field theories available to date.

For massless particles such as photons and approximately energetic astroparticles~(neutrinos, in particular, $\kn,E\gg m_{\nu}$ and $\kn\simeq E$), from an astrophysical source~(say, a GRB) at redshift $z$, the time difference due to~(\ref{eq:mdr}) can be written upon taking into account the cosmic expansion as~\cite{Jacob:2008bw,Ellis:2002in}
\be\la{eq:gtd}
\D t_{\textrm{LV}}(E,z)=(1+z)s_{n}\frac{\K_{n}(\D E,z)}{\Elvn^{n}},
\ee
where $n=1$ corresponds to linear dependence of the particle energy~(the case of particular interest to us here), $\D E^{n}$ is the energy difference between particles of high energies $E_{h}$ and those with low energies $E_{l}$, $\D E^{n}=E_{h}^{n}-E_{l}^{n}$, measured in the observer reference frame, and
\be\la{eq:lvf}
\K_{n}(\D E,z)\coloneqq\frac{1+n}{2}\frac{\D E^{n}}{(1+z)H_{0}}\int_{0}^{z}\sd\z\frac{(1+\z)^{n}}{\sqrt{\Omega_{\Lambda}+\Omega_{m}(1+\z)^{3}}},
\ee
is the~($n$th-order) LV factor, where $H_{0}$ is the present Hubble parameter, and $\Omega_{\Lambda}$, $\Omega_{m}$ are the cosmological parameters evaluated today. In this type of study, $E_{l}$~(of order $\Or(1-100~\textrm{keV})$) is negligible, as it is significantly smaller than $E_{h}$ which lies in the range of GeV to multi-TeV for cosmic photons and up to PeV for neutrinos, hence in~(\ref{eq:gtd}), $\D E\simeq E_{h}$. The authors of~\cite{Jacob:2008bw} first derived the formula in $\Lambda$CDM model, and it was further verified in Finsler space-times where MDRs is introduced consistently~\cite{Zhu:2022blp,Zhu:2023mps}. Although the analysis is insensitive to cosmology~\cite{Biesiada:2009zz,Pan:2015cqa}, nevertheless, there are cosmological-model independent approaches also, see for instance~\cite{Zou:2017ksd,Zhang:2018ooj,Pan:2020zbl}.

In~\cite{Ellis:1999sd,Ellis:2002in}, the authors reported early studies for this type of test, prior to the launch of the high-precision Fermi $\gamma$-ray Space Telescope~(FGST) and later in~\cite{Ellis:2005sjy}, as we now come to introduce, a robust strategy to do collective analysis over the photon data from different observations has been developed. The Large Area Telescope~(LAT) on board the FGST~\cite{Fermi-LAT:2009ihh} discovered that high-energy photons arrived $\Or(0.1-10~\textrm{s})$ later relative to the low-energy ones~\cite{Fermi-LAT:2009owx,FermiGBMLAT:2009nfe}, which might present potential evidence for subluminal LV expected by the above D-particle space-time foam models. However, the determination of the possible LV-induced time lag from observational data is not a trivial task. Apart from conventional astrophysical explanation of such effects~(i.e., the matter in the Universe may also cause dispersion effects and associated delays in the propagation of light~\cite{Latimer:2013rja}), photons are in general not radiated simultaneously at the source. So, there is an intrinsic source-induced lag $\Delta t_{\textrm{int}}$ which is not known, due to our imperfect knowledge of the radiation mechanism of GRBs. It has been suggested, though, in~\cite{Ellis:2005sjy,Shao:2009bv} that the contributions from source effects can be disentangled from that of QG effects, if one can achieve a statistical survey, combining data from multiple GRB sources at different redshifts. The basic idea is that, the LV time delay~(\ref{eq:gtd}) which is a gravitational medium effect as in the string D-foam models~(c.f.~(\ref{eq:dft}), above) accumulates with the propagation distance but intrinsic time difference will not. So, the observed arrival time lag of the particle is $\D t_{\textrm{obs}}=\D t_{\textrm{LV}}+(1+z)\D t_{\textrm{int}}$, that is what one can obtain from data, $\D t_{\textrm{obs}}=t_{h}-t_{l}$, where $t_{h(l)}$ is the arrival time of the high~(low) energy event. Since $\D t_{\textrm{int}}$ is unknown, in the analysis it is useful to reexpress this formula as
\be\la{eq:obsd}
\frac{\D t_{\textrm{obs}}}{1+z}=\D t_{\textrm{int}}+s_{n}\frac{\K_{n}}{\Elvn}.
\ee
The suffixes $n$ in the above formulae shall be omitted in the case of the linear form correction to the particle velocity, i.e., the $n=1$ case, that is the situation largely favored by the current data~(see below).

As we can see from~(\ref{eq:obsd}) if the energy-dependent speed variation of GRB photons really exist, there would be a linear correlation between $\D t_{\textrm{obs}}/(1+z)\eqqcolon\Y$ and $\K_{n}$. Photons with same intrinsic emission time would fall onto a same line in the $\Y$-$\K_{n}$ plot, and then one can extract the value of $\Elvn/s_{n}$ from the slope of that line and determine $\D t_{\textrm{int}}$ of them as the $\Y$-intercept. Initially, as in the seminal work~\cite{Ellis:2005sjy} of this area, $\D t_{\textrm{int}}$ is taken as a constant, $\D t_{\textrm{int}}(\textrm{const.})$~\cite{Shao:2009bv,Zhang:2014wpb,Xu:2016zxi,Xu:2016zsa,Amelino-Camelia:2016ohi,Amelino-Camelia:2017zva,Xu:2018ien,Liu:2018qrg,Chen:2019avc,Zhu:2021pml,Zhu:2021wtw,Zhu:2022usw}, whilst we note that it appears essential to invoke a more general form for intrinsic lags~\cite{Song:2025qej,Song:2025apj,Song:2025akr} with, say, a dependence on the photon energy in GRB source frame~\cite{Song:2024and}. We shall come back later on this point.

The situation changes significantly when one extends this type of LV study to neutrinos, because unlike photons, neutrinos can escape from dense astrophysical environments. UHE cosmic neutrinos can then reach the Earth, serving as an ideal portal to reveal the tiny QG effects~(if the latter exist)~\cite{Jacob:2006gn,Amelino-Camelia:2009imt,Amelino-Camelia:2015nqa}. Indeed, the energies of extragalactic neutrinos detected by IceCube neutrino observatory range from several TeVs to a few PeVs, so the time delay or advance of PeV neutrinos can be up to a few months. Although till this day, for the UHE neutrino events~\cite{IceCube:2013cdw,IceCube:2013low,IceCube:2014stg} observed, the collaboration announces no significant connection with GRBs in a short temporal window~(of a few hundred seconds), nevertheless from the above consideration the authors of~\cite{Amelino-Camelia:2016fuh} first associated IceCube 60--500~TeV ``shower'' events in an expanded time range of 3 days with GRB candidates, based on the best correlation together with direction and time criteria~\cite{Amelino-Camelia:2016fuh,Amelino-Camelia:2016wpo,Amelino-Camelia:2016ohi,Amelino-Camelia:2017zva,Huang:2018ham}.

With the above-described strategy of analysis, a positive signal that may be compatible with LV was exposed in time-of-flight photon delays and from the probable associations of IceCube neutrinos with GRBs~\cite{Shao:2009bv,Zhang:2014wpb,Xu:2016zxi,Xu:2016zsa,Amelino-Camelia:2017zva,Xu:2018ien,Liu:2018qrg,Chen:2019avc,Zhu:2021pml,Zhu:2021wtw,Zhu:2022usw,Song:2024and,Song:2025qej,Song:2025apj,Song:2025akr,Li:2020uef,He:2022gyk,Amelino-Camelia:2016ohi,Huang:2018ham,Huang:2019etr,Huang:2022xto}:

\bi
\im \underline{Light-Speed Variation from GRB/AGN Photons?}\vspace{0.15cm}

The time-of-flight of GRB $\gamma$-rays is the clearest test to search for the minuscule effects of LV of the photon sector as GRBs are intense flashes within milliseconds due to collapsing massive stars at distant parts of the Universe. It provides thus the most convincing results. In~\cite{Shao:2009bv,Zhang:2014wpb}, high-energy photons of multi-GeV energies from different GRBs of the Fermi telescope were analyzed collectively, and the later work~\cite{Xu:2016zxi,Xu:2016zsa} used the first main peak in the low-energy light curve rather than the trigger time of the Gamma-ray Burst Monitor~(GBM)~\cite{Meegan:2009qu} on board the FGST as the low-energy characteristic time, $t_{l}$. Via testing arrival time differences between energetic photons with the corresponding low-energy photon emissions for Fermi-LAT GRBs with known redshifts, it was found with surprise that, eight events from five long bursts fall on an inclined line~(which we refer as the mainline afterward in later studies) in the $\Y$-$\K_{n=1}$ plot, and this regularity revealed for the time lags between photons of different energies gained strong support from the 51.9~GeV event of GRB~160509A, that falls exactly on the same mainline~\cite{Xu:2016zsa}. This regularity indicates a \emph{linearly-suppressed} light-speed variation which is \emph{subluminal}, $s=+1$, and of the form\footnote{It has been demonstrated explicitly~\cite{Xu:2016zxi} that as compared to quadratic-order correction of the light speed, the linear LV scenario is more favored by data with stronger regularities.}
\be\la{eq:slsv}
v_{\gamma}(E)=1-E/\Elvg,\quad\textrm{with}~\Elvg\sim 3\times 10^{17}~\textrm{GeV},
\ee
i.e., of the type expected in D-foam models~\cite{Li:2021gah}. Some progress has been made in~\cite{Xu:2018ien,Liu:2018qrg,Chen:2019avc,Zhu:2021pml,Zhu:2021wtw,Zhu:2022usw} on testing the effect carefully from analysis of high-energy photons~(in the GeV and TeV range) observed by Fermi and other instruments, such as the Major Atmospheric Gamma-ray Imaging \v Cerenkov~(MAGIC) telescope and LHAASO. It is worth noticing that this estimate on $\Elvg$ is comparable to plenty of bounds set by similar TOF studies of GRB $\gamma$-ray data~\cite{Ellis:2002in,Ellis:2005sjy,Bolmont:2006kk,Lamon:2007wr,Bernardini:2017tzu,Ellis:2018lca,Bartlett:2021olb}. AGNs~\cite{Biller:1998hg,MAGIC:2007etg,HESS:2019rhe,Li:2020uef}, and pulsars~\cite{Kaaret:1999ve,Zitzer:2013gka}, were also utilized to detect in-vacuo light-speed dispersion with this method, and the results are consistent with~(\ref{eq:slsv}), including the constraint~\cite{Li:2023kgi} from recent observations of $\gamma$-rays up to 20~TeV from the Vela pulsar~\cite{HESS:2023sxo}.\vspace{0.05cm}

In fact, just about a year before the Fermi measurements, in July 2006, the High Energy Stereoscopic System~(H.E.S.S.) observed a time lag $\sim 20$~s~\cite{HESS:2008irp} between energy bands of 200--800~GeV and $>800$~GeV, during a flare of the blazar PKS~2155-304, leading to a slighter tighter lower bound for $\Elvg$, which seems incompatible with Equation~(\ref{eq:slsv}). However, as argued in the later study~\cite{Li:2020uef}, it is possible to accommodate the findings of H.E.S.S. with the light-speed variation from GRB photons, provided that the latter can be trusted. It was also shown that the $\Or(1~\textrm{min})$ delays of the TeV photons of AGN Markarian~(Mrk) 501~\cite{MAGIC:2007etg,Albert:2007zd}, at the MAGIC, can be fitted with the same LV scenario as the early observation of Mrk 421, for which no distinct lag was found between light curves of different energy bands~\cite{Biller:1998hg,Shao:2009bv}.\footnote{In contrast to GRB photon lag tests, these studies were carried out based upon the assumption $\D t_{\textrm{int}}=0$, as the time is not yet ripe for a joint analysis of AGNs with similar features but different redshifts to reveal the source effects, due to the scarcity of data available~\cite{Amelino-Camelia:2009imt,Ellis:2008gg}.}\vspace{0.05cm}

We also notice, of course, the discrepancies in the limits to $\Elvg$ in several other studies, as compared to the above findings. The primary factor contributing to the conflicts lies in the additional premises, about the intrinsic timing for photons of different energies, that these studies commonly made as they only study individual $\gamma$-rays from a \emph{single} source. It is in this way that trans-Planckian lower limits on the scale are inferred from, for instance the $\Or(0.1~\textrm{s})$ delays of the $\sim 31$~GeV photon from the extremely short burst GRB~090510~\cite{Xiao:2009xe,Bolmont:2010np,Nemiroff:2011fk,Vasileiou:2013vra,Vasileiou:2015wja}, observed by Fermi~\cite{FermiGBMLAT:2009nfe}, and/or the MAGIC 1.07~TeV event of the GRB~190114C~\cite{MAGIC:2020egb,MAGIC:2019lau}. In contrast, there is no such assumption in a statistical approach to TOF, with $\D t_{\textrm{int}}$ derived solely from data fitting.\vspace{0.05cm}

The global fits of 14 multi-GeV photons from 8 Fermi GRBs directly pointed to negative $\Y$-intercepts~(see Figure 2 of~\cite{Xu:2016zsa}) and for the mainline photons $\D t_{\textrm{int}}\approx -10.7$~s, thereby implying that, at sources, energetic photons are emitted prior to softer ones. The short GRB~090510 event falls in this case on a parallel line lower than the mainline indicating the same $\Elvg$ as~(\ref{eq:slsv}). Such a prediction on a \emph{preburst} stage for emission of high-energy photons~(before the prompt phase of GRBs) has been further examined in~\cite{Chen:2019avc,Zhu:2021pml,Zhu:2021wtw,Huang:2019etr,Song:2024and} with favorable signals, supporting the cooling nature of GRBs. In one of these recent works~\cite{Song:2024and}, the authors reanalyze the same dataset of Fermi by incorporating a possible source energy-$E_{\textrm{sr}}$-dependent term for intrinsic time delay of photons,
\be\la{eq:edid}
\D t_{\textrm{int}}=\alpha E_{h,\textrm{sr}}+\D t_{\textrm{int}}(\textrm{const.}),\quad E_{h,\textrm{sr}}= (1+z)E_{h},
\ee
where $\alpha$ is a proportional factor and we discard contributions from low-energy photons for reasons mentioned before, and they also imply consistent results with the previous works~\cite{Xu:2016zxi,Xu:2016zsa}. Still, as the whole analysis yielding such encouraging outcomes remains preliminary at present, evidently, we have to see whether they will withstand, as more data are accrued. Fortunately, the operation of LHAASO that has already kicked off a new era in $\gamma$-ray astronomy due to their discovery of PeV photons~\cite{LHAASO:2021gok}, can help in this purpose. They lately measured very-high energy~(VHE) $\gamma$-ray emission~($E\gtrsim 100$~GeV) beyond 10~TeV from a GRB.
\bi
\im[$\divideontimes$] \underline{LHAASO Observations of GRB~221009A}~\cite{Huang:2022grb,LHAASO:2023kyg,LHAASO:2023lkv}\vspace{0.15cm}

LHAASO is a new-generation air shower detector array in China, aiming to study the Universe at VHEs/UHEs in cosmic and $\gamma$-rays~\cite{LHAASO:2019qtb,Cao:2021vqs}. It includes a surface Water \v Cerenkov Detector Array~(WCDA), a Kilometer Square Array~(KM2A), and a Wide Field-of-view \v Cerenkov Telescope Array~(WFCTA), covering a total area of $1.36~\textrm{km}^{2}$. On 9 October 2022, LHAASO observed the unprecedentedly brightest GRB~221009A~($z\simeq 0.15$). After the GBM trigger~\cite{Lesage:2023vvj,Veres:2022grb}, WCDA captured over 64,000 events in the 0.2--7~TeV energy range~\cite{LHAASO:2023kyg} while KM2A registered~\cite{LHAASO:2023lkv} more than 140 events with energies above 3~TeV, recording the highest photon statistics in the TeV band ever from a GRB. The greatest GRB photon energy, ever observed, is about 99.3~GeV with Fermi-LAT~\cite{Pillera:2022grb,Bissaldi:2022grb} and about 251~TeV by Carpet-2~\cite{Dzhappuev:2022grb}. LHAASO initially reported photons up to 18~TeV~\cite{Huang:2022grb} while later refined analysis in~\cite{LHAASO:2023lkv} reduces the value to roughly 12.2~TeV. Named brightest-of-all-time~(BOAT) this GRB was also registered by other space missions~\cite{Dichiara:2022grb,Krimm:2022grb,Ursi:2022grb,Piano:2022grb,Gotz:2022grb,Xiao:2022grb,Frederiks:2023bxg,Frederiks:2022grb,Lapshov:2022grb,Ripa:2023ssj,Ripa:2022grb,Mitchell:2022grb,Liu:2022grb,Insight-HXMT:2023aof,Duan:2022grb} and ground-based detectors like Carpet-3, which lately reports even higher energy~($\sim 300$~TeV)~\cite{Dzhappuev:2025ase}. These characteristics render this signal a unique opportunity to detect LV in the photon sector.
\ei
By adopting the newly proposed model for intrinsic timing~(\ref{eq:edid}) in order to take proper account of the complex behaviors of GRB source effects, the very recent studies~\cite{Song:2025apj,Song:2025qej,Song:2025akr} analyze three exceptional events---the MAGIC 1.07~TeV event of GRB~190114C, the FGST 99.3~GeV photon and LHAASO 12.2~TeV photon of BOAT---alongside the prior Fermi-LAT data with a Bayesian parameter estimation method~\cite{Song:2024and}. Intriguingly, the results support the same physical scenario for photons covering an energy band from a few tens of GeVs up to a dozen of TeVs with $\alpha,\D t_{\textrm{int}}<0$ and $\Elvg\sim 3\times 10^{17}$~GeV. The direct evidence in support of the preburst stage of BOAT is presented in~\cite{Liu:2024qbt}. It is this feature that elucidates why these observations differ from those stringent limits in the literature~\cite{LHAASO:2024lub,Piran:2023xfg,Yang:2023kjq}. As shown in~\cite{Song:2025apj}, the functional form of $\D t_{\textrm{int}}$~(\ref{eq:edid}) is selected by data, devoid of any biased assumptions. If one is tempted to have higher-order terms, say, $\D t_{\textrm{int}}=\D t_{\textrm{int}}(\textrm{const.})+\alpha E_{\textrm{sr}}+\beta E_{\textrm{sr}}^{2}+\gamma E_{\textrm{sr}}^{3}$, or a redshift dependence of the factors, like $\alpha(z)=\alpha+\alpha_{1}(z-\bar{z})+\alpha_{2}(z-\bar{z})^{2}$~(where $\bar{z}$ is the average redshift of all 17 events), the fits show explicitly that $\beta,\gamma\rightarrow 0$, $\alpha_{1},\alpha_{2}\sim 0$, respectively. It is remarkable that the null-hypothesis of dispersion-free vacuum~(i.e., the constant light-speed, $v_{\gamma}=1$) is rejected at a significance level of $3.1~\sigma$~\cite{Song:2025apj}. Evidently, these findings indicate, therefore, for photons traveling through cosmic space, the presence of light-speed modifications, of the type expected~\cite{Li:2021gah} to be encountered in models of QG-induced space-time foam, coming from brane/string theory~\cite{Li:2021eza,Li:2022sgs,Li:2023wlo,Li:2024crc}.
\im \underline{Neutrino-Speed Variation from IceCube Events?}\vspace{0.15cm}

This strategy has been not only applied to GRB photons and also to GRB-neutrinos~\cite{Amelino-Camelia:2016fuh,Amelino-Camelia:2016wpo,Amelino-Camelia:2016ohi,Amelino-Camelia:2017zva,Huang:2018ham,Huang:2019etr,Huang:2022xto}~(and the combination of the two~\cite{Amelino-Camelia:2016ohi}). By ``GRB-neutrinos'' here we mean candidate GRB neutrinos, suggested via such kind of analyses, from the plausible associations of IceCube neutrino events with GRBs. IceCube has observed plenty of UHE neutrinos of energies beyond 30~TeV, including a couple of PeV events~\cite{IceCube:2013cdw,IceCube:2013low,IceCube:2014stg,IceCube:2016uab}. While GRBs are considered one class of potential sources of these cosmic neutrinos~(as in the fireball model of GRBs~\cite{Waxman:1997ti}, for instance), the collaboration only reported correlations of GRBs with neutrinos of lower energies around 1~TeV~(we term them ``near-TeV'' events), in a close time window, mentioned previously, deducing that they are compatible with the atmospheric backgrounds~\cite{IceCube:2014jkq,IceCube:2016ipa,IceCube:2017amx}.\vspace{0.05cm}

In~\cite{Amelino-Camelia:2016ohi}, the authors made the breakthrough to identify coincident detections of neutrinos and photons from the same GRB, via expanding the temporal window between the neutrinos and associated GRBs to a few days. This is reasonable provided the time differences due to LV corrections could be extended by the ultra-high energies and the cosmologically large propagation distances of neutrinos in a quantum space-time. For the nine shower neutrinos with energies from 60~TeV to 500~TeV, at the IceCube, they revealed~\cite{Amelino-Camelia:2016fuh,Amelino-Camelia:2016wpo,Amelino-Camelia:2016ohi} roughly compatible speed variation features as exposed in analyses of GRB photons from associating these neutrinos with several GRBs. It was soon noticed that four IceCube PeV neutrinos~(three shower events~\cite{IceCube:2013cdw,IceCube:2013low,IceCube:2014stg} plus a ``track'' event~\cite{IceCube:2016uab}) are associated with GRBs within a time range of three months~\cite{Huang:2018ham}. These four PeV-scale neutrinos~(with energies up to the highest at 2.6~PeV) fall on the same straight line for the events in the multi-TeV range, implying a \emph{linearly} energy-dependent speed variation of GRB-neutrinos: 
\be\la{eq:nsv}
v_{\nu}(E)=1\mp E/\Elvnu,\quad\textrm{with}~\Elvnu\simeq 6.4\times 10^{17}~\textrm{GeV},
\ee
The estimate of the scale $\Elvnu$ for cosmic neutrinos is consistent with other TOF bounds available today, coming from, say, neutrino pulses from supernovae~\cite{Ellis:2008fc,Ellis:2011uk}, or, other energetic events from IceCube, e.g.,~\cite{Wang:2016lne}.\footnote{We note that the limit of $\Elvnu>0.01\Ep$ therein was derived from the association~\cite{Kadler:2016ygj} between one of the PeV events~(\#35) with the blazar PKS B1424-418. But, even if the event came from this blazar instead of the GRB suggested in~\cite{Huang:2018ham} their result is still compatible with the value of~(\ref{eq:nsv}).} It is also compatible with those limits~\cite{Ellis:2018ogq,Laha:2018hsh,Wei:2018ajw} from multimessenger observations of blazar TXS~0506+056 coincident with a 290~TeV tracklike neutrino~\cite{IceCube:2018cha,IceCube:2018dnn}.\vspace{0.05cm}

Rather intriguingly, contrary to the GRB photons, there are \emph{both} ``late''~($\D t_{\textrm{obs}}>0$) and ``early''~($\D t_{\textrm{obs}}<0$) neutrinos, also referred to as time delay, $\D t_{\textrm{LV}}>0$~(c.f. $s=+1$), and, advance $\D t_{\textrm{LV}}<0$~($s=-1$) events for TeV and PeV scale neutrinos,\footnote{For near-TeV events~\cite{Huang:2019etr} it is not the case, as their observed lags are competitive in magnitude to intrinsic ones.} as the intrinsic lag $\D t_{\textrm{int}}$~(of order $\Or(1~\textrm{hr})$ determined by data) is too short to make a difference compared to LV time lags~(a few days to months). This can be explained by different propagation properties between neutrinos and antineutrinos provided the IceCube detector cannot tell the chiralities for the neutrinos~(except for Glashow resonance~\cite{Glashow:1960zz} events, like the one lately registered~\cite{IceCube:2021rpz}). Then, either neutrinos, or antineutrinos, are superluminal, and the other ones are subluminal, indicating CPT violation of the neutrino sector, or, an asymmetry between matter and antimatter~\cite{Zhang:2018otj}. Nonetheless, we still need to seek additional support for such findings. It was soon showed that 12 near-TeV northern hemisphere track events from IceCube fall on the same line~\cite{Huang:2019etr}, and that another 3 or more multi-TeV track neutrinos~(above 30~TeV) can also be associated with GRBs under the same Lorentz violation scenario~\cite{Huang:2022xto}.\vspace{0.05cm}

We lately noted~\cite{Li:2024crc}, however, that all these track events being probably GRB neutrinos with retarded travel times~(i.e., delay events) might point to a stronger signature for the subluminal aspect of cosmic neutrino LV, if compared to the 60~TeV to 2~PeV regularity, just revisited~(\ref{eq:nsv})~\cite{Amelino-Camelia:2016fuh,Amelino-Camelia:2016wpo,Amelino-Camelia:2016ohi,Amelino-Camelia:2017zva,Huang:2018ham}. In fact, IceCube collaboration recently performed significant revision of the directional information of the neutrinos~\cite{IceCube:2020wum}, and a reanalysis~\cite{Amelino-Camelia:2022pja} of the recalibrated data of events $\gtrsim 60$~TeV investigated whether~(and, if so, how) this affects the above LV/CPTV features in cosmic neutrinos. It was found that, for events subject to propagation time delays, the signal is even stronger than deduced with the previous incorrect estimates of the directions, while there seems less encouragement for the QG effects speeding up neutrinos~(or, antineutrinos)~(given the whopping probability of accidentally finding that feature).\vspace{0.05cm}

Savvy readers may recognize that this scenario is consistent with the stringy space-time foam models where neutrinos/antineutrinos can only be slowed down~\cite{Li:2024crc,Ellis:1999sf,Ellis:2008gg}, as discussed in the previous section. However, in our opinion, support from the presently available data for early neutrinos, undergoing acceleration by the effects of LV, has yet not completely evaporated, not even models able to fit such a picture, especially the QG backgrounds of stochastic D-foam~\cite{Li:2022sgs,Li:2023wlo} which, as we have seen, entail CPT-violating vacuum refraction effects for neutrinos. It has been found~\cite{Zhang:2018otj} that, interpreting these time advance events with the CPT-odd feature of the linear LV modification within EFT framework faces difficulties, because, such superluminal neutrinos quickly lose energy via, say, $\nu\rightarrow\nu ee^{+}$. In contrast, such an issue does not arise in our stochastic D-particle foam model where, as already mentioned, violations of the energy conservation during such processes enable the practically stable propagation of these~(anti)neutrinos, hence the evasion of the stringent constraints~(in Section~\ref{sec:6}). This is an advantage of the brane D-foam approach, where consistent QG descriptions for the findings of neutrino-speed variation in~\cite{Amelino-Camelia:2016ohi,Amelino-Camelia:2017zva,Huang:2018ham,Huang:2019etr,Huang:2022xto} can be developed~\cite{Li:2023wlo,Li:2024crc}.\vspace{0.05cm}

Evidence in favor of the preburst stage of GRBs has also been found from such neutrino analyses. For instance, in~\cite{Huang:2019etr}, an intrinsic time difference which, again, is negative and about $-300$~s was obtained alongside~(\ref{eq:nsv}) from data fits. Combined with LV studies on GRB photons~\cite{Shao:2009bv,Zhang:2014wpb,Xu:2016zxi,Xu:2016zsa,Amelino-Camelia:2017zva,Xu:2018ien,Liu:2018qrg,Chen:2019avc,Zhu:2021pml,Zhu:2021wtw,Zhu:2022usw,Song:2024and,Song:2025qej,Song:2025apj,Song:2025akr}, the results suggest a physical scenario of intrinsic properties that may be in support of the GRB fireball model~\cite{Waxman:1997ti}: about a few hundred seconds before the outburst of low-energy electromagnetic signals at the source, a pre-neutrino burst first happens, and because of the big time lags or leads for the velocity variation of UHE neutrinos, we can observe them even a few months later or before the arrival of $\gamma$-ray signals; then, high-energy photons come out prior to the subsequent prompt low-energy photons, but they travel slower than the latter due to light-speed variation. So some of them can be observed after low-energy lights, as in Fermi GRBs~\cite{Fermi-LAT:2009owx,FermiGBMLAT:2009nfe}, while others should arrive before the trigger if the preburst stage exists, as is indeed the case in the LHAASO observation of BOAT~\cite{Liu:2024qbt}.\vspace{0.05cm}

Despite finding intriguing preliminary indications of QG, as advocated here, we have to stress that the results should be treated with care, since they still rely on very limited data. There is no reason to jump to any conclusions, also because at current stage, none of these ``GRB-neutrinos'' has been ascertained as being neutrinos from GRBs: some of the neutrino--GRB associations can occur just accidentally~(without any linear LV effect). But if such associations become numerous enough, as new data are available, it would, evidently, be meaningful. Most recently, the KM3NeT detectors observed the highest-energy cosmic neutrino ever registered~\cite{KM3NeT:2025npi,KM3NeT:2025aps}. And when associated with a GRB suggested by the authors of~\cite{Amelino-Camelia:2025lqn}, this 220~PeV neutrino provides a bound for the subluminal LV scale: $\Elvnu\sim (4-10)\times 10^{17}$~GeV, which is compatible with the priorly determined value in~(\ref{eq:nsv}). There still remain other potential~(GRB) sources~\cite{Wang:2025lgn}, with $\Elvnu$ varying up to an order of magnitude around $\Or(10^{17}~\textrm{GeV})$. We expect that more refined multimessenger analyses to be developed in the future may allow to further check the revealed regularity.
\ei
Whether or not these preliminary statistical signatures of LV imprints on the propagation of particles withstand further experimental tests, analyses revisited here are signaling the start of a more mature phase of QG phenomenology. Particularly, during the last few decades of fundamental physics, results even more encouraging than these have gradually faded away, and we are therefore well prepared to see these cosmic photons and neutrinos have that fate or, if any of the features here contemplated find greater support, as more data are accrued. Nonetheless, findings of a strong correlation between energy and time-of-arrival~(\ref{eq:gtd})~(where $n=1$)---a correlation of just the type expected in the aforementioned D-particle space-time foam models~(c.f.~(\ref{eq:dft})) due to induced index of refraction \emph{in vacuo}---have motivated possible interpretations of the effects by such models~\cite{Li:2021gah,Li:2021eza,Li:2022sgs,Li:2023wlo,Li:2024crc}. However, if the latter have a chance of being true, they must respect all other complementary astrophysical constraints for LV, available to date. As we mentioned earlier, it seems that, at present, only this kind of string theory models of quantum foam, where only photons and neutrinos are susceptible to foam effects, could stand up to this chance.

A positive signal of LV in photon travel-time lags at the scale of a few $10^{17}$~GeV would defy birefringence constraints, to be briefly reviewed in the next section, at least in the SME framework~(\ref{eq:d5a}) but this never poses a dilemma if it can be attributed to~(subluminal) photon propagation in D-brane foam, which leads to the absence of birefringence~\cite{Mavromatos:2010pk,Li:2021gah,Li:2021eza}. Then, for a uniform D-foam density, at least for late epochs of the Universe's history, say, $z\leq 10$~(the GRB farthest from us, $z\sim 8$), an order of magnitude,
\be\label{eq:cdf}
\frac{\Ms}{\gs n_{\Dp}(z)}\sim 10^{17}~\textrm{GeV},\quad n_{\Dp}(z<10)\simeq\textrm{const.},
\ee
determined for the effective QG mass, by comparison with the above findings of the energy-dependent speed variations of GRB photons and neutrinos, is sufficient to make the D-foam explanation of such latter effects viable. On the other hand, with this estimate, such models can also accommodate the rather impressive results both for the CPTV neutrino scenario, and for the subluminal neutrino picture with enhanced signals~\cite{Amelino-Camelia:2022pja} in the corrected IceCube data~\cite{Li:2023wlo,Li:2024crc}.

Key observation here is that, different scenarios of LV yield different phenomenological predictions: while delays in the time-of-flight of ultrarelativistic particle could be a common signature, other constraints, from modifications or, appearance of thresholds of reactions, to birefringence constraints, do not apply in the same form. And many more measurements, at various cosmological distances~(redshifts) are needed, before we falsify a model, such as the string D-foam scenario(s), for instance, espoused here. With new data, that Fermi, LHAASO and, IceCube, KM3NeT, or future Tropical Deep-sea Neutrino Telescope (TRIDENT)~\cite{Ye:2023dch}, will continue to report but still for some time to come, this will hopefully become possible in the foreseeable future.

\section{Complementary Tests with Astroparticles}\label{sec:6}

The sensitivity of the purely kinematical time-of-flight observations to the Planck-scale physics~(\ref{eq:cdf}), at least for a linearly suppressed effect in high-energy astrophysics, calls for an immediate confrontation with other sensitive probes of nontrivial optical properties of QG medium. Indeed these latter measurements offer alternative~(and independent) ways to test Lorentz invariance, and, imply already very stringent constraints for specific types of LV, as we now come to discuss, mainly focusing on observational limits in the photon--electron of the QED and neutrino sectors. We shall be brief in our description of these complementary astrophysical tests on QG-inspired MDRs, to avoid large digression from our main point of this review which is to discuss LV situations where string theory predictions can be falsified by~(astro)particle physics processes and experiments.

One of the most discussed linear-order LV modifications to photon dispersion relations, for example in EFT LV dimension-5 QED~(\ref{eq:d5a}), or loop QG model of~\cite{Gambini:1998it}, entails a dependence of light speed on photon polarization, i.e., birefringence, as the ground state of such theories breaks reflexion symmetry~(parity). The fact that group velocities of photons of same energy but opposite helicities, denoted by $\pm$, differ slightly in such models,
\be\la{eq:hdls}
v_{\gamma(\pm)}=1\mp\omega/\Elvg^{\pm},
\ee
indicates that the polarization vector of a linearly polarized light~(of frequency $\omega$) rotates, in the course of wave propagation over a cosmological distance, through the angle~\cite{Laurent:2011he,Toma:2012xa},
\be\la{eq:bra}
\lvert\D\theta_{\textrm{LV}}(\omega,z)\rvert\simeq (1+z)\frac{\K(\omega^{2},z)}{2\Elvg^{\pm}},\quad\textrm{for}~n=1~\textrm{LV},
\ee
where we plugged in the LV-sign~(or, helicity) factor $s=\pm 1$, and the scale $\Elv^{\pm}$ describes the broken degree of parity/CPT~(not to be confused with $\Elv$, which is related with \emph{either} the subluminal~($s=+1$) or superluminal~($s=-1$) scenario). Polarimetric observations of light, from a plethora of astrophysical objects including the Crab Nebula~\cite{Maccione:2008tq}~(e.g.,~\cite{Gleiser:2001rm,Jacobson:2003bn,Fan:2006xuj,Stecker:2011ps,Shao:2011uc,Gotz:2014vza,Lin:2016xwj,Wei:2019nhm}; for recent bounds, see Table 1 of~\cite{Li:2021eza}, and for reviews~\cite{Jacobson:2004rj,Mattingly:2005re,Jacobson:2005bg,Liberati:2009pf,Liberati:2013xla,Wei:2021ite}) rule out this phenomenon unless it occurs at a scale~(way) beyond the Planck mass, $\Elvg^{\pm}>(10^{7}-10^{16})\times\Ep$~(or, one expects the order of suppression of the effect, $n>1$).

An important comment arises at this stage concerning the meaning of a LV scale higher than the Planck mass. First, we note that such results can be translated also into constraints in the LV SME model~(\ref{eq:d5a}), indicating $\xi$ far below its natural order $\Or(1)$, which means that extra suppression of the LV effects has to be invoked. Similarly, this may be considered a strong indication that certain~($n=1$ birefringent in this case) types of QG-inspired LV are excluded on naturalness grounds, because there is no mechanism to account for $\Elv\gg\Ep$ within these contexts. Nonetheless, we should mention that what one calls a natural scale of QG is highly model dependent and it depends, for instance, in string models of D-foam, on a combination of model parameters~(c.f.~(\ref{eq:cdf})), hence, any possible trans-Planckian values of the scale, if required by future GRB arrival time test, can be easily satisfied with more dilute foam populations.

Consequently, if quantized space-time is indeed causing energy-dependent corrections to the speed of particles at the first order, birefringent LV scenarios with a helicity-dependent speed of light~(\ref{eq:hdls}) where \emph{both} $s=+1$ and $s=-1$ are permissible will be strongly disfavored. On the other hand, it is clear from our analysis~(in Section~\ref{sec:4}, above) and that of~\cite{Li:2021gah,Li:2021eza,Li:2023wlo,Li:2024crc,Mavromatos:2010pk}, there is no microscopic model dependence of the induced deformations of the light propagation, other than the subluminal nature of the D-foam induced refractive index and the associated \emph{lack} of birefringence, namely the blindness of foam to photon helicities. This is an important feature, which allows the phenomenological findings of light-speed variation~(\ref{eq:slsv}) to survive naturally the stringent constraints obtained from vacuum LV birefringence tests as we have explained in closing the previous section.

An interesting phenomenology of threshold reactions has been introduced, besides the above-cited TOF and birefringence studies if one adopts a phenomenological framework of LV, justifying the use of deformed particle dispersion relations. However, not all these tests are similarly robust for the underlying QG physical framework, because, the cumulative effects exclusively use the form of the energy-dependent speed variations~(\ref{eq:mdr}), while testing LV corrections in particle reactions would depend on the use of such MDRs, and the usual assumption of strict energy--momentum conservation. Such premises stem from the validity of local EFT description of QG effects on particles with $E\ll\Ep$. Thus, the upshot of this is the Hamiltonian dynamics, such that the velocity $\delta_{{\cal I}}$~(\ref{eq:mdr}) can be derived from the following MDR~(albeit not necessarily regarding particular QG (D-foam) model mentioned earlier):
\be\la{eq:gmdr}
E_{{\cal I}}^{2}-\pv_{{\cal I}}^{2}+(sE\pv^{2}/\Elv)_{{\cal I}}\simeq m_{{\cal I}}^{2},
\ee
where the suffix ${\cal I}$ denotes the particle types, ${\cal I}=\gamma$, $e^{\mp}$, or, $\nu/\bar{\nu}$. With the standard laws of energy--momentum conservation this can lead to peculiar~(threshold) reactions beyond SM, characterized by the following features~\cite{Mattingly:2002ba,Jacobson:2002hd}:
\bi
\im Existing thresholds of reactions can shift and upper thresholds (i.e., maximal incoming momenta below which the process is allowed to occur) can appear.
\im New, anomalous~(normally forbidden) channels can happen.
\im Threshold configurations still correspond to head-on collisions with parallel outgoing particles.
\ei

There are three major classes of complementary astrophysical limits, coming from such effects of the photon, electron and neutrino, respectively, to be considered in any attempt to interpret the speed variation~(\ref{eq:slsv}),~(\ref{eq:nsv}), or more general $\gamma$-ray/neutrino astrophysics results, in terms of QG-induced anomalies in particle dispersion.
\bi
\im[(1)] \underline{Implications or Bounds from Energetic $\gamma$-Ray Annihilation and Self Decay.}\vspace{0.15cm}
\bi
\im[$\divideontimes$] It has been well established that in SR, pair production, $\gamma\gamma_{b}\rightarrow ee^{+}$, plays a crucial role in making our Universe opaque to energetic $\gamma$-rays as these are annihilated in this way by very-low energy cosmic background lights~($\gamma_{b}$) such as the ones of the IR extragalactic background light~(EBL), or, cosmic microwave background~(CMB) radiation, that populates the Universe today, as a relic from the Big-Bang era. Let us call $\omega_{b}$ the~(low) energy of the target photon $\gamma_{b}$; then, the threshold in SR for the annihilation occurs with the $\gamma$-ray photon having energy, $E_{\gamma}=m_{e}^{2}/\omega_{b}\simeq 411$~TeV, if the soft photon is from CMB~\cite{Li:2021cdz,Li:2021tcw}, or about 261~GeV for EBL absorption~\cite{Li:2022vgq}. This results in an attenuation of the measured photon flux at VHEs and UHEs, as such reactions prevent photons above these~(lower) thresholds from propagating vast distances in cosmic space.\vspace{0.05cm}

In the presence of subluminal-photon MDRs, like the ones used in the aforementioned QG-interpretation of the effect~(\ref{eq:slsv}), the threshold is in general altered, such that CMB~(or, EBL) may be more transparent to high-energy $\gamma$-rays~(compared to the Lorentz invariant case). Indeed, it has been found that, the threshold, for the soft photon to absorb a high-energy $\gamma$-ray~\cite{Jacob:2008gj,Shao:2010wk,Li:2021tcw},\footnote{The electron/positron LV effect is temporarily ignored in obtaining~(\ref{eq:spth}) from the threshold theorem~\cite{Mattingly:2002ba}~(since, it is absent in D-foam models and indeed severely constrained by experiments; see below.)}
\be\la{eq:spth}
\omega_{b,\textrm{th}}(E_{\gamma})=\frac{m_{e}^{2}}{E_{\gamma}}+\frac{1}{4}\Bigl(\frac{E_{\gamma}}{\Elvg}\Bigr)E_{\gamma},\quad\textrm{for}~s_{\gamma}=+1,
\ee
has a minimum, which can be attained at the critical energy, $E_{\gamma,\textrm{cr}}=(2m_{e}^{2}\Elvg)^{1/3}$, then that, for $E_{\gamma}>E_{\gamma,\textrm{cr}}$, with the increase of $\gamma$-ray energy, $\omega_{b,\textrm{th}}$ also increases. As the number density of background lights decreases
with energy, there are less soft photons to annihilate with energetic $\gamma$-rays above $E_{\gamma,\textrm{cr}}$, therefore, a recovery of the high-energy photon flux is expected. Detailed analysis by considering instead the thresholds of high-energy photon, $E_{\gamma,\textrm{th}}$, exposes three distinct LV phenomena from the kinematics~(threshold) equation, at the leading order~\cite{Li:2021cdz},
\be\la{eq:ppte}
\ell_{\gamma}=f(k)\coloneqq\frac{4\omega_{b}}{k^{2}}-\frac{4m_{e}^{2}}{k^{3}},\quad\textrm{for}~k>0,
\ee
where $\ell_{\gamma}\coloneqq s_{\gamma}/\Elvg$, and we assumed their 4-momenta $p_{\gamma}=(E_{\gamma},0,0,k)$, $p_{\gamma_{b}}=(\omega_{b},0,0,-\omega_{b})$, where $E_{\gamma}$ and $k=\lvert\pv_{\gamma}\rvert$ obey~(\ref{eq:gmdr}), while any LV of soft photon is neglected as $\omega_{b}\ll E_{\gamma}$. With the fact that $f(k)$ has only one zero $k=m_{e}^{2}/\omega_{b}$ that is just the threshold of SR, while that the maximum of $f$ occurs at $k_{\textrm{cr}}=3m_{e}^{2}/(2\omega_{b})$, where $\ell_{\gamma,\textrm{cr}}\equiv\max f$, one can sort out whether $\ell_{\gamma}=f(k)$ has a solution, that is, a threshold, $k_{\textrm{th}}$, and whether $k_{\textrm{th}}$~(if more than one) are lower or upper ones leading to different threshold behaviors~\cite{Li:2021cdz}:
\bi
\im[\textbf{Case~a}]$\qquad$ If $\ell_{\gamma}>\ell_{\gamma,\textrm{cr}}=16\omega_{b}^{3}/(27m_{e}^{4})$, even a lower threshold does not exist, then subluminal photons with $\Elvg<\ell_{\gamma,\textrm{cr}}^{-1}\simeq 4.5\times 10^{23}$~GeV, e.g.,~(\ref{eq:slsv}), cannot be absorbed by background photons of energy $\omega_{b}$~(here, taken as CMB photons, as an illustration, with the mean energy $\omega_{b}\simeq 6.35\times 10^{-4}$~eV). As $\gamma\gamma_{b}\rightarrow ee^{+}$ is kinematically forbidden, the \emph{optical transparency} of CMB to cosmic photons of high energies is expected. 
\im[\textbf{Case~b}]$\qquad$ If $0<\ell_{\gamma}<\ell_{\gamma,\textrm{cr}}$, an upper threshold $k_{\textrm{th}}^{>}$~\cite{Kluzniak:1999qq,Mattingly:2002ba} can develop for which the process does not occur. Thus, when interacting with the $\gamma_{b}$ background, a subluminal photon can create an $ee^{+}$ pair only if $E_{\gamma}\in [E_{\gamma}(k_{\textrm{th}}^{<}),E_{\gamma}(k_{\textrm{th}}^{>})]$, with the lower threshold $k_{\textrm{th}}^{<}>m_{e}^{2}/\omega_{b}$. This implies the \emph{reappearance} of photons at energies above $E_{\gamma}(k_{>})$.
\im[\textbf{Case~c}]$\qquad$ If $\ell_{\gamma}<0$, there is only one solution for~(\ref{eq:ppte}), and this $k_{\textrm{th}}^{<}$ is smaller than expected in SR and QED, namely a \emph{reduction} of the lower threshold. However, as photons are superluminal, the $\gamma$-decay process dominates, where it results in a cutoff and no photons are observable above the decay threshold~(c.f.~(\ref{eq:pdt}), below), so we shall not dwell much on this case.
\ei
For systematic studies on such threshold anomalies in, say, $n>1$ LV, see~\cite{Mattingly:2002ba,Jacobson:2002hd}. We note that such phenomena were first advanced to resolve the so-called ``TeV-$\gamma$ paradox''~(c.f. for 20~TeV photons striking IR background lights, the experimental indication of an excess in the spectra of Mrk 421 and Mrk 501)~\cite{Kluzniak:1999qq,Kifune:1999ex,Protheroe:2000hp,Amelino-Camelia:2000ono,Amelino-Camelia:2000bxx} and more refined studies on LV impacts on EBL opacity were carried out with observations of especially TeV blazars~\cite{Biteau:2015xpa,Tavecchio:2015rfa,Horns:2016vfv,Franceschini:2017iwq,Abdalla:2018sxi,Lang:2018yog}.\vspace{0.05cm}

Recently, an important part of the interest on this is due to the first measurements of PeV cosmic photons by LHAASO~\cite{LHAASO:2021gok}. While the fact, that without LV, photons with energies above the threshold 411~TeV should be depleted during their travel to Earth by striking the CMB radiation, may be pointing to new physics beyond SM~\cite{Li:2021tcw,Li:2021duv} such PeV events originating in the nearby Cygnus can still be observed upon taking into account the absorption lengths~(free paths) of these photons~\cite{Li:2021cdz,Ling:2021ytu}. However, if multi-TeV and PeV photons can be associated with extragalactic origins according to the strategy depicted in~\cite{Li:2021cdz,Li:2022wxc} they may serve as convincing signals for the subluminal photon LV features as \textbf{Case a} and \textbf{Case b} both indicate. To quantitively study the attenuation by EBL in the form of $\e^{-\tau_{\gamma}}$, the optical depth $\tau_{\gamma}(E_{\gamma},z)$, which quantifies the dimming of the spectrum of the source, at redshift $z$, is introduced~\cite{Biteau:2015xpa,Li:2022vgq},
\be\la{eq:od}
\tau_{\gamma}(E_{\gamma},z)=\int_{0}^{z}\sd\z\frac{\sd t}{\sd\z}\int_{-1}^{1}\sd\mu_{\theta}\frac{1-\mu_{\theta}}{2}\int_{\varpi_{b,\textrm{th}}}^{\infty}\sd\varpi_{b}n_{\omega}(\varpi_{b},\z)\sigma_{\gamma\gamma}({\cal E}_{\gamma},\varpi_{b},\mu_{\theta}),
\ee
where $\mu_{\theta}\equiv\cos\theta$, with the angle, $\theta$, between two interacting photons, $\theta\in [-\pi,\pi]$, $n_{\omega}\coloneqq\pd n/\pd\varpi_{b}$ is the EBL photon number density, $\varpi_{b,\textrm{th}}=\omega_{b,\textrm{th}}({\cal E}_{\gamma},\mu_{\theta})$, ${\cal E}_{\gamma}=(1+\z)E_{\gamma}$ and $\sigma_{\gamma\gamma}$ the cross section of the process. For a flat Universe, the differential of time is, $\sd t=[(1+\z)H_{0}\sqrt{\Omega_{m}(1+\z)^{3}+\Omega_{\Lambda}}]^{-1}\sd\z$. Recently, threshold anomalies were used, widely~\cite{Li:2022vgq,Li:2022wxc,Baktash:2022gnf,Finke:2022swf,Zheng:2022ooe,Li:2023rgc,Li:2023rhj}, to interpret possible excesses of VHE $\gamma$-rays of GRB~221009A, which may pose a challenge to standard physics as it requires more transparency in intergalactic space than traditionally expected~\cite{LHAASO:2023lkv}. We notice that while for WCDA photons $\lesssim 10$~TeV~\cite{LHAASO:2023kyg}, the suppression due to EBLs still allows them to arrive on Earth, the 18~TeV photon, as initially reported by KM2A~\cite{Huang:2022grb}, for which the flux is suppressed by at least $10^{-8}$~\cite{Li:2022vgq,Li:2022wxc}, signals an excess, that is very likely to contradict the SR. The authors of~\cite{Li:2023rhj} showed that subluminal photon LV with $\Elvg\lesssim 0.1\times\Ep$ provides a viable explanation for this signal after correcting for EBL absorption~\cite{LHAASO:2023lkv}.\footnote{Whether that 18~TeV photon clashes with conventional physics~\cite{Zhao:2022wjg} and necessitates LV effect~\cite{Galanti:2022chk,LHAASO:2023lkv} remains a matter of debate. However, taking at face value the highest-energy Carpet event from GRB~221009A now at 300~TeV~\cite{Dzhappuev:2025ase} would unequivocally imply an evidence for anomalies.}~(Alternatively, one may consider an axion origin of this event~(e.g.,~\cite{Zhang:2022zbm,Wang:2023okw,Galanti:2022chk}).) Despite uncertainties of EBL models~\cite{Li:2023rgc} we still find that, this primary suggestion is consistent with the aforementioned findings from GRB time lag studies, with $\Elvg$ found via such threshold anomaly analyses, very near the reduced Planck mass, of order $10^{18}$~GeV, which notably is the order of $\Ms$ in traditional string theories.
\im[$\divideontimes$] While, normally, the process for a single photon to create an $ee^{+}$ pair~(spontaneous decay) is forbidden by SR it may be allowed by the presence of LV. Indeed, setting $\omega_{b}=0$ in~(\ref{eq:ppte}) yields a wild estimate of the threshold, consistent with that derived from the threshold theorem~\cite{Mattingly:2002ba},
\be\la{eq:pdt}
k_{\textrm{th}}\simeq\Bigl(-\frac{4m_{e}^{2}\Elvg}{s_{\gamma}}\Bigr)^{1/3},\quad\textrm{for}~E_{\gamma}\gg m_{e},
\ee
which allows the process if $s_{\gamma}=-1$~(c.f. \emph{superluminal} photons). It was shown~\cite{Martinez-Huerta:2016odc,Martinez-Huerta:2017ntv} that the rate goes like $k^{2}/\Elvg$ above threshold, and is of the order $(10~\textrm{ns})^{-1}$ for a 10~TeV photon~\cite{Jacobson:2005bg,Mattingly:2005re}. Hence, $\gamma\rightarrow ee^{+}$, once kinematically allowed, is quite efficient~\cite{Jacobson:2002hd} and would lead to a sharp cutoff in the spectrum of $\gamma$-rays at UHEs beyond which no particle should reach Earth. This justify the use of the threshold values~(\ref{eq:pdt}) to constrain superluminal LV in photons, so the explicit limit from any observation of high-energy cosmic $\gamma$-ray is $\Elvg\gtrsim E_{\gamma}^{3}/(4m_{e}^{2})$.\vspace{0.05cm}

This phenomenon has been extensively studied in the contexts of general MDRs and effective field theories pertaining to the SME~(see, for instance in~\cite{Coleman:1997xq,Klinkhamer:2008ky,Mattingly:2005re,Jacobson:2005bg,Liberati:2009pf,Liberati:2013xla}). Earlier tests were reported via considering the stability of $\sim 20$~TeV photons from the Crab Nebula~\cite{Coleman:1997xq} and adopting later in~\cite{Stecker:2001vb,Stecker:2003pw} the detection of events up to 50~TeV~\cite{CANGAROO:1997aci}, while the advent of new data from the High-Energy-Gamma-Ray Astronomy~(HEGRA) telescopes~\cite{HEGRA:2004tpc}, and the observations of the supernova remnant~(SNR) RX~J1713.7-3946 by H.E.S.S.~\cite{Aharonian:2006ws}, allowed more restrictive limits~\cite{Schreck:2013paa,Klinkhamer:2008ky,Martinez-Huerta:2016azo} due to the nonobservation of the effect. A tighter result comes from $\gamma$-rays around 100~TeV detected by High Altitude Water \v Cerenkov~(HAWC) telescopes~\cite{HAWC:2019gui} where, combining the Crab and three other sources, the exclusion limit $\Elvg>2.22\times 10^{22}$~GeV was inferred at $2\sigma$.\vspace{0.05cm}

A significant improvement was presented very recently in~\cite{Li:2021tcw,Li:2021duv}~(and in, e.g.,~\cite{Chen:2021hen,LHAASO:2021opi}) with the result that exceeds the sensitivity of the aforementioned experiments to such superluminal LV photon features by $\gtrsim (1-4)$ orders of magnitude. These studies take advantage of the breakthrough made by LHAASO to detect the UHE $\gamma$-ray emission of 12 Galactic PeVatrons~\cite{LHAASO:2021gok}. A crude estimate about the strongest limit one can obtain from this dataset was deduced as $\Elvg>2.7\times 10^{24}$~GeV~\cite{Li:2021tcw} by inserting $E_{\gamma}\simeq 1.42$~PeV of the extremely high-energy single event in~(\ref{eq:pdt}), i.e.,
\be\la{eq:spsc}
\Elvg\gtrsim 9.57\times 10^{32}~\textrm{eV}\Bigl(\frac{E_{\gamma}}{1~\textrm{PeV}}\Bigr)^{3},\quad\textrm{for}~s_{\gamma}=-1,
\ee
while, later, the collaboration article~\cite{LHAASO:2021opi} complements the study of~\cite{Li:2021tcw}, reporting similar~(and more robust) results based on the lack of sharp spectrum cutoffs in two of their sources. The most recent findings of LHAASO on the highest-energy photon signal of energy $\sim 2.5$~PeV, ever observed by a human, from the Cygnus bubble~\cite{LHAASO:2023uhj}, further tighten the above bound by a factor of $\sim 5$~\cite{Li:2025ste}. Such results severely restrict LV theories allowing $\gamma$-decays such as dimension-5~(CPT-odd) LV operators~(\ref{eq:d5a}), within the SME, where the induced photon MDR~(\ref{eq:d5pdr}) is nothing but an analogy to the phenomenological relation~(\ref{eq:gmdr}) in the context by the replacement $(s_{\gamma}/\Elvg)\mapsto (\mp\, 2\xi/\Mp)$. The resulting constraint, $\xi<\Or(10^{-6}-10^{-7})$~\cite{Li:2021tcw,Li:2025ste}, is, however, still about 10 orders of magnitude weaker than the most stringent limit to date on this birefringent SME coefficient $\xi$ from polarimetry of GRBs~\cite{Wei:2019nhm}. As becomes clear from our discussion in Section~\ref{sec:3}, the subluminal photons, which are also characterized by $\xi$ in the model, thus suffer from same strong constraints, that are several orders of magnitude smaller than the value required to reproduce the GRB photon lags, should the effect~(\ref{eq:slsv}) be attributed predominantly to photon propagation in a LV vacuum, indicating the failure of EFT approach in explaining speed
variation of cosmic photons~\cite{Li:2021tcw}.\vspace{0.05cm}

In our above discussion of photon decay, any LV effects for electrons~(or, positrons) were neglected for both theoretical and observational considerations. We note, for completeness, that in a phenomenological context, one may allow its presence and see what constraint can we infer from current data. In fact, if we set $\Elvg=\infty$~(c.f. assuming Lorentz invariant photons), then, a direct test of subluminal LV of the electron/positron sector rises from the search of $\gamma$-decay, which is kinematically permitted once $E_{\gamma}\geq (8m_{e}^{2}\Elve/s_{e})^{1/3}$; so, this threshold is meaningful, only if the outgoing charged fermions possess \emph{subluminal} LV dispersions, $s_{e}=+1$~\cite{Jacobson:2002hd,Jacobson:2001tu}. The observed 2.5~PeV photon suggests, then, that the scale should be higher than $10^{5}\Mp$~\cite{Li:2025ste} that is indeed very stringent. In case one assumes LV present in \emph{both} fermions and photons, i.e., having nonzero $\ell_{e}\coloneqq s_{e}/\Elve$ and $\ell_{\gamma}$, the constraint depends on both parameters, forming excluded regions in parameter space. This was done thoroughly in~\cite{He:2022jdl,He:2023ydr}, using 1.4 PeV as the highest photon energy, prior to the report of the 2.5 PeV event.
\ei
It should be emphasized that in the aforementioned analysis of threshold anomaly, we assumed that energy--momentum conservation law remains intact, despite the MDRs for the photons. However, this may not be the case, as the local EFT formulation of the propagation effects~(namely, a low-energy representation of the QG-medium \emph{dispersive effects} with higher-derivative \emph{local operators} in flat space-time Lagrangians) underlying such a premise, may break down. This is indeed the case of certain~(stochastic) models of D-foam mentioned earlier, where the fluctuations of the space-time or other defects of gravitational nature play the role of an external environment, resulting in energy violations in $\gamma\gamma_{b}\rightarrow ee^{+}$. So that, whether threshold anomalies exist in such models is still an open question,\footnote{A primary study of $\gamma\gamma\rightarrow ee^{+}$ was carried out in~\cite{Ellis:2000sf}, focusing on an earlier treatment of a recoiling D-particle. Unfortunately, that class of model differs from what we study here~(in Section~\ref{sec:4})~(within which the behavior of the process is up for debate), hence we cannot make direct use of those results.} while for the stretched-string framework~\cite{Ellis:2008gg,Ellis:2009vq}, it is clear that the kinematics of the process is identical with that in conventional QED~\cite{Ellis:2010he}. On the other hand, as light propagation in string D-particle foam is necessarily subluminal, photons are \emph{stable}~(i.e., do \emph{not} decay), obviating any potential problems from $\gamma$-decays. This allows for a consistent interpretation of light-speed variation with those $\gamma$-decay limits within such models~\cite{Li:2021gah,Li:2021eza}.
\im[(2)] \underline{Synchrotron and \v Cerenkov Constraints to~(Electron) LV Dispersions}\vspace{0.15cm}

At this point we would like to describe that, apart from any theoretical motivation from D-foam models, in current astrophysical measurements there is also no encouragement for the hypothesis of QG effects modifying the dispersion relations of charged leptons, primarily electrons and positrons, which are the ones we consider below. Indeed, for the subluminal type of LV corrections on their in-vacuo dispersions, mentioned above, besides photon decay, these would affect the synchrotron radiation of distant galaxies, such as the Crab Nebula~\cite{Jacobson:2002ye,Ellis:2003ua,Ellis:2003sd,Ellis:2003if,Maccione:2007yc}. In fact, the most severe of all known limits to electron/positron LV arise from observations of this object, which is a SNR that was witnessed in 1054~A.D. and lies only about 1.9~kpc from Earth. It is characterized by the energetic QED processes, exhibiting a well-studied broad spectrum, with characteristic double peak of the synchro-self-Compton~(SSC) mechanism~\cite{Atoyan:1996kqd,Meyer:2010tta}, i.e., the leptonic acceleration mechanism, whereby the \emph{same electron~(positron)} responsible for \emph{synchrotron} emission from the radio band of the Crab also undergoes \emph{inverse-Compton~(IC) scattering} to produce high-energy $\gamma$-rays above 1~GeV~\cite{Li:2022ugz}.\vspace{0.05cm}

In both Lorentz symmetric and LV cases, electrons in a magnetic field ${\bf B}$ at core regions of the nebula(e) follow helical orbits transverse to the direction of ${\bf B}$. The so-accelerated electrons emit synchrotron radiation with a spectrum that sharply cuts off at a critical frequency~\cite{Jacobson:2002ye,Ellis:2003sd},
\be\la{eq:srcf}
\omega_{\textrm{cr}}(E_{e})=\frac{3}{2}e\lvert {\bf B}\rvert\frac{\gamma_{v}^{3}(E_{e})}{E_{e}},
\ee
where $e$ is the electric charge, and $\gamma_{v}=(1-v_{e}^{2}(E_{e}))^{-1/2}$, with $v_{e}$, the electron~(group) velocity. In standard relativistic physics, $E_{e}=\gamma_{v}m_{e}$, so, $\omega_{\textrm{cr}}=(3/2)eBE_{e}^{2}/m_{e}^{3}$~(where $B\coloneqq\lvert {\bf B}\rvert$) grows with $E_{e}$ without bound. This is affected by LV~\cite{Jacobson:2002ye,Jacobson:2003bn}; for an electron with $s_{e}=+1$,\footnote{Although we also have LV in the photon sector, the emitted frequencies of the synchrotron radiation are much lower than the energy of the source particles~(electrons/positrons). Effectively, LV for the synchrotron photons can be ignored here~\cite{Jacobson:2005bg}.} it has a maximal attainable velocity strictly less than the low-energy speed of light $\delta_{e}\simeq -(m_{e}^{2}/2E_{e}^{2}+E_{e}/\Elve)<0$, hence there is a maximum synchrotron frequency it can produce, regardless of its energy~\cite{Jacobson:2004rj,Jacobson:2005bg,Liberati:2009pf,Liberati:2012jf,Liberati:2013xla}:
\be\la{eq:crfm}
\max\omega_{\textrm{cr}}=\frac{eB}{m_{e}}\Bigl(\frac{m_{e}}{\Elve}\Bigr)^{-2/3}\C,\quad\textrm{for}~s_{e}=+1,
\ee
where $\C=5^{5/6}/(9\root 3 \of 2)\simeq 0.34$, a small constant. This maximal frequency is attained at $E_{e}=(2m_{e}^{2}\Elve/5)^{1/3}$. The authors of~\cite{Jacobson:2002ye}, using the largest estimated value 0.6~mG for $B$, jointly with the fact that the Crab synchrotron emission has been found to extend at least up to energies of about 100~MeV, obtained from $\max\omega_{\textrm{cr}}\geq 0.1$~GeV a stringent constraint $\Elve>1.7\times 10^{26}$~GeV. While more refined analysis later in~\cite{Maccione:2007yc} weakened it by $(1-2)$ orders of magnitude, now one has this more robust bound on the relevant subluminal electron LV scale at $2\sigma$.\vspace{0.05cm}

If $s_{e}=-1$ instead, the electron speed can exceed the constant speed of light, at which point $\gamma_{v}(E_{e})$ diverges. This corresponds to the threshold of the soft \v Cerenkov emission. In the presence of LV, \v Cerenkov effect of leptons, $e\rightarrow e\gamma$, can occur in a vacuum; above thresholds, the rate of the so-caused energy loss scales as $E_{e}^{3}/\Elve$, which implies that a 10~TeV electron would emit a significant fraction of its energy in $10^{-9}$~ns, short enough that constraints derived solely from threshold analysis alone are again reliable~\cite{Jacobson:2005bg,Li:2025ste}. As the rate of the electron decays via $e\rightarrow e\gamma$ is orders of magnitude greater than the IC scattering rate, any electron known to propagate~(stably, and therefore, appropriate for the production of VHE IC $\gamma$-rays) in the Crab must lie below the \v Cerenkov thresholds. This yields the so-called IC \v Cerenkov constraint.\footnote{The competing synchrotron energy loss is irrelevant for this constraint. For $B\sim 1$~mG~(as in SNRs), the rate due to synchrotron emission is 40 orders of magnitude smaller than that of $e\rightarrow e\gamma$~\cite{Jacobson:2002hd}.} If the emitted \v Cerenkov photon is soft enough so that its LV can be neglected, $\gamma_{v}$ reaches infinity at a finite energy, which amounts to $v_{e}\approx 1$~\cite{Li:2022ugz}, i.e.,
\be\la{eq:sct}
E_{e}\simeq\Bigl(-\frac{m_{e}^{2}\Elve}{2s_{e}}\Bigr)^{1/3},\quad\textrm{for}~E_{e}\gg m_{e}.
\ee
So, indeed, it is meaningful only if $s_{e}=-1$ and constraints on the possible \emph{superluminal} LV scale for electrons/positrons can be extracted via matching with observations. The fact that electrons are stable against the soft \v Cerenkov effect at energies up to at least 450~TeV inferred from the 450~TeV $\gamma$-rays~\cite{TibetASg:2019ivi} arriving on Earth from the Crab Nebula has implied an improvement of about two orders of magnitude~\cite{Li:2022ugz} over limits~\cite{Jacobson:2003bn} from the previous energies, in the range 50--80~TeV, which are necessary to explain the earlier photon observations of up to 75~TeV~\cite{CANGAROO:1997aci,HEGRA:2004tpc}. Then, handling~(\ref{eq:sct}), the result is $\Elve>2E_{e}^{3}/m_{e}^{2}\sim 7\times 10^{23}$~GeV.\vspace{0.05cm}

Very recently, LHAASO-KM2A announced~\cite{LHAASO:2021cbz} for the first time the Crab IC spectrum past 1 PeV, and the highest photon energy registered is about 1.12~PeV. Detailed studies on the SSC model show that the energy of the parent electron is around 2.3~PeV~\cite{LHAASO:2021cbz,Li:2022ugz}. This allows us to report in~\cite{Li:2022ugz}, for the superluminal electron dispersions, the strongest constraint ever obtained, $\Elve>9.4\times 10^{25}$~GeV from the absence of the soft \v Cerenkov threshold~(\ref{eq:sct}) up to such a high energy scale, i.e.,
\be\la{eq:sesc}
\Elve\gtrsim 7.66\times 10^{33}~\textrm{eV}\Bigl(\frac{E_{e}}{1~\textrm{PeV}}\Bigr)^{3},\quad\textrm{for}~s_{e}=-1.
\ee
This result improves previous IC \v Cerenkov constraints~\cite{Jacobson:2001tu,Jacobson:2002hd,Jacobson:2003bn} by $10^{4}$ times, and further restricts theories that entail such MDRs, such as the SME QED modified by a dimension-5 LV term~(\ref{eq:d5a}), where values of $\eta_{\textrm{R/L}}>10^{-7}$ are excluded as a result. There is also the possibility that $e\rightarrow e\gamma$ involves emission of a high-energy photon~\cite{Jacobson:2002hd,Konopka:2002tt}. For a hard emitted photon, the \v Cerenkov threshold~(in the absence of electron/positron LV for simplicity, that is also the case of particular relevance for us) occurs at $E_{e,\textrm{th}}\simeq (4m_{e}^{2}\Elvg/s_{\gamma})^{1/3}$; therefore, the lack of such hard  \v Cerenkov effects implies strict limits for the photon LV scale in the \emph{subluminal} case~($s_{\gamma}=+1$), and the 2.3~PeV long living electrons require this scale to be higher than $1.2\times 10^{25}$~GeV~\cite{He:2022jdl,He:2023ydr}, which exceeds the sensitivity of GRB time lag studies~(c.f.~(\ref{eq:slsv})) to such subluminal photon LV properties by seven orders of magnitude.\vspace{0.05cm}
 
It was also shown, in~\cite{He:2022jdl,He:2023ydr}, that if one takes into account MDRs for both photons and electrons, the parameter space, that is still permitted by LHAASO 1.4~PeV photon~\cite{LHAASO:2021gok} and~(indirectly observed) 2.3~PeV electron~\cite{LHAASO:2021cbz}, lies in a narrow region in the proximity of $\ell_{\gamma}=\ell_{e}\geq 0$, which indicates that the LV scale of photons should be of comparable orders of magnitude as that of electrons if the subluminal LV effects really act on them. For subluminal LV electrons, as we have mentioned, there are strong constraints for the relevant scale coming from synchrotron radiation studies, $\Elve\gtrsim 10^{24}$~GeV~(derived at $2\sigma$~\cite{Maccione:2007yc}). So, from the result of~\cite{He:2023ydr}, one may conclude that in the case of linear $\Mp$ suppression of the subluminal QG-induced LV for photons, of interest for exotic~(QG) interpretation of light-speed variation effect~(\ref{eq:slsv}), scales with size $\Elvg\simeq\Or(\Elve)<10^{24}$~GeV are ruled out. However, this is not the case~\cite{Li:2025ste}, and the lack of comparison with a meaningful QG theory represents a severe limitation of the analysis based on a purely phenomenological approach to MDRs.\vspace{0.05cm}

And we note at this point that, such results only \emph{excludes the possibility} that the findings suggested from GRB photons are due to a LV effect that acts \emph{universally} among photons and electrons. However, these cannot exclude the anomalous photon propagation with linear energy dependence in models where the Lorentz-violating QG foam is transparent to electrons~\cite{Ellis:2003ua,Ellis:2003sd,Ellis:2003if,Li:2022ugz}, as in the D-foam case. As shown in~\cite{Li:2025ste}, electrons emit \emph{no} \v Cerenkov radiation, despite traveling faster than photons~(of comparable energies), which experience vacuum refraction in this approach. Hence if the effect of subluminal light-speed variation could finally be attributed partly or wholly to this type of stringy space-time foam, as put forward in~\cite{Li:2021gah,Li:2021eza,Li:2022sgs,Li:2023wlo,Li:2024crc}, then, the IC~(soft) \v Cerenkov bounds, as well as the impressive hard \v Cerenkov constraint for photons considered in~\cite{He:2022jdl,He:2023ydr}, are no longer applicable. The synchrotron radiation measurements of the Crab Nebula are also consistent with such models, provided from such observations one could basically rule out subluminal LV effect in electron dispersion while in D-foam scenarios, LV is exactly absent for charged particles.~(Therefore, we are justified in assuming Lorentz invariance in the electron sector for our discussion below).
\im[(3)] \underline{Pair Creation and Constraints on Neutrino Velocities}\vspace{0.15cm}

To interpret the preliminary statistical finding, that some of the high-energy neutrinos observed by IceCube might be GRB neutrinos whose travel times were affected by LV, and therefore by microscopic properties of space-time~(foam), especially the result that appears to support a CPT-violating propagation for cosmic neutrinos~\cite{Amelino-Camelia:2016ohi,Huang:2018ham,Huang:2019etr,Huang:2022xto}, one must deal with the apparent problem brought about by neutrino superluminality. As we have seen, superluminal LV typically allows new particle decay processes, that are forbidden in Lorentz invariant case. Specifically, superluminal neutrinos could exhibit distinct energy-loss channels~\cite{Cohen:2011hx}, such as the \v Cerenkov radiation, $\nu\rightarrow\nu\gamma$, neutrino splitting $\nu\rightarrow\nu\nu\bar{\nu}$ and bremsstrahlung~(i.e., pair creation/emission) effect $\nu\rightarrow\nu ee^{+}$, but, as mentioned, this assertion relies on strict energy--momentum conservation that is applicable in the framework of low-energy EFT of LV.\vspace{0.05cm}
 
In the wake of IceCube cosmic-neutrino detections, stringent constraints on LV within the neutrino sector have been imposed, since neutrinos of high energies would struggle to propagate large distances to Earth if LV effects were allowing these peculiar channels particularly $\nu\rightarrow\nu ee^{+}$, which dominates neutrino-energy loss, to occur. For the pair emission, since the neutrino transforms almost into an $ee^{+}$ pair, each with $1/2$ energy of the original, the threshold is given by
\be\la{eq:pcth}
E_{\nu,\textrm{th}}\simeq 2m_{e}\Bigl(-\,s_{\nu}\frac{E_{\nu}}{\Elvnu}\Bigr)^{-1/2}=\frac{2m_{e}}{\sqrt{\delta_{\nu}}}.
\ee
It is valid for generic models of~(linear) LVs with $s_{\nu}=-1$~(c.f. \emph{superluminal} neutrinos). From the decay width~\cite{Cohen:2011hx,Huo:2011ve} which tells the free path of the neutrino, one can derive constraints on $\delta_{\nu}$, which is required to produce at least the propagated distance to the source, as long as the process is permitted due to threshold effects~(\ref{eq:pcth}). Observations of $\gtrsim 100$~TeV diffuse neutrinos, as well as the very existence of~(above) PeV neutrinos provide already bounds as tight as~\cite{Borriello:2013ala,Stecker:2013jfa,Diaz:2013wia,Wang:2020tej} $\delta_{\nu}<10^{-18}$ to $\sim 5.6\times 10^{-19}$ for a positive $\delta_{\nu}$. This corresponds to $\Elvnu\geq (10^{3}-10^{5})\Ep$, which seems to exclude any possibility of superluminal neutrino propagation with linear Planck-scale suppression, as in~(\ref{eq:nsv}). Recent analyses~\cite{KM3NeT:2025mfl,Satunin:2025uui} exploiting the KM3NeT 220~PeV neutrino event~\cite{KM3NeT:2025npi} further improved the sensitivity to such superluminal Lorentz-violating neutrino velocities to the level $\delta_{\nu}<\Or(10^{-22}-10^{-23})$. The resulting constraint translated into a linear LV scale is now stronger by as much as thirteen orders of magnitude than the finding~(\ref{eq:nsv}) inferred from IceCube-neutrino--GRB associations by the presence of LV, should the candidate GRB-neutrinos from~\cite{Amelino-Camelia:2016ohi,Amelino-Camelia:2017zva,Huang:2018ham,Huang:2019etr,Huang:2022xto} and~\cite{Amelino-Camelia:2016fuh,Amelino-Camelia:2016wpo} be ascertained eventually as being signals from the respective sources.\vspace{0.05cm}

Although the detection of the UHE neutrino events leads to neutrino~(meta)stability~(or long lifetime) which merely means that the relevant decay channels are not operational, it does impose strong constraints for some field-theoretic models of LV~\cite{Zhang:2018otj,Carmona:2023mzs}. And, as shown in~\cite{Zhang:2018otj}, the attempt to interpret the aforementioned CPTV neutrino-speed variation with such models faces inevitably challenge due to the constraints. However, the situation changes significantly if one goes beyond the EFT approach. The above results simply cannot be used to invalidate the superluminal LV~(and CPTV) effects on neutrino propagation, particularly when neutrinos~(or, antineutrinos) do not undergo decay despite moving faster the constant speed of light. That seems kind of impossible, maybe counterintuitive, but in Section~\ref{sec:4}, we actually encountered already an example of the latter, that of a stochastically recoiling D-foam and the associated superluminal antineutrinos in the model~\cite{Li:2023wlo,Li:2022sgs}.\vspace{0.05cm}

For previous constraints that apply to antineutrinos for the CP-conjugated channels as understood, the argument that the model evades such restrictions on $\delta_{\bar{\nu}}$~(or $\Elvnu$ for $\delta_{\bar{\nu}}>0$) by considerations of instability of the particles has its roots in the foam-induced deformation of energy--momentum conservation, which leads to a modification of the kinematics of the superluminal $\bar{\nu}$-decay process. Let us remind that in the stochastic D-foam background, for $\delta_{\bar{\nu}}$ deduced from~(\ref{eq:sfndr}), the process $\bar{\nu}\rightarrow\bar{\nu}ee^{+}$ is kinematically allowed provided that~\cite{Li:2022sgs},
\be\la{eq:spcth}
E_{\bar{\nu}}\gtrsim\frac{2m_{e}}{\sqrt{\varDelta}},\quad\textrm{where}~\varDelta\coloneqq\Bigl(1-\frac{4\varsigma_{I}}{\fd_{\Dp}^{2}}\Bigr)\delta_{\bar{\nu}},
\ee
or equivalently, $E_{\bar{\nu}}\geq E_{\bar{\nu},\textrm{th}}\equiv (4m_{e}^{2}{\cal E}_{\ast})^{1/3}$ from Equation~(\ref{eq:fpcth}); the new energy scale, ${\cal E}_{\ast}$, introduced here for $\bar{\nu}\rightarrow\bar{\nu}ee^{+}$ is given by
\be\la{eq:sfpcs}
{\cal E}_{\ast}=\frac{\Elvnu}{1-4\varsigma_{I}\Elvnu/\Md},
\ee
where $\delta_{\bar{\nu}}\simeq\fd_{\Dp}^{2}E_{\bar{\nu},\textrm{th}}/\Md=E_{\bar{\nu},\textrm{th}}/\Elvnu$ at threshold was used. We again observe that the channel is open only if $\fd_{\Dp}^{2}>4\varsigma_{I}$. If this were the case, one can envision a situation where $\fd_{\Dp}\approx 2\sqrt{\varsigma_{I}}$ then, the threshold would be pushed to a very high energy scale of, for instance order PeV, so that $\bar{\nu}\rightarrow\bar{\nu}ee^{+}$ never produces a depletion of PeV antineutrino fluxes during their~(superluminal) propagation.\vspace{0.05cm}

Nonetheless, as discussed in~\cite{Li:2023wlo,Li:2022sgs}, current observations can provide~(merely, though) bounds on a \emph{combined quantity} of the fundamental parameters of the foam. For
example, the detection of the 2~PeV event, IceCube \#35~\cite{IceCube:2013cdw,IceCube:2013low,IceCube:2014stg}, a time advance event~(according to the analysis~\cite{Huang:2018ham}) and hence probably a superluminal antineutrino induced by D-foam effects, implies a limit $(\fd_{\Dp}^{2}-4\varsigma_{I})\lesssim 8.4\times 10^{-8}$ for $\fd_{\Dp}>2\sqrt{\varsigma_{I}}$. While a more refined analysis should extract the constraint from the interaction length $\sim\Gamma^{-1}$, we obtained this result based on consideration that the threshold~(\ref{eq:spcth})~(or~(\ref{eq:fpcth})) should be of $\Or(\textrm{PeV})$ for observing events of such an energy.\footnote{Lacking, at present, a complete theory of the matter--D-foam interactions, we did not consider the rate $\Gamma$ above threshold. Nonetheless~\cite{Li:2023wlo,Li:2022sgs}, to first order in $1/\Md$, the matrix element for any process agrees with that of special-relativistic QFT, while the recoil effects of D-particles modify the \emph{kinematics} of the field-theoretic result.} Thus, if the neutrino-speed variation and the consequent Lorentz/CPT violation for cosmic neutrinos as in~\cite{Amelino-Camelia:2016ohi,Amelino-Camelia:2017zva,Huang:2018ham,Huang:2019etr,Huang:2022xto} were due to this type of foam, then, the tight bounds so far cast by means of neutrino $ee^{+}$ pair creation, e.g.,~\cite{Borriello:2013ala,Stecker:2013jfa,Diaz:2013wia,Wang:2020tej,KM3NeT:2025mfl,Satunin:2025uui}, are actually imposed on $\varDelta$ or on the scale ${\cal E}_{\ast}$ in the D-foam case; these are naturally evaded if ${\cal E}_{\ast}\leq 0$ or easily satisfied in case of ${\cal E}_{\ast}>0$ with an assignment for the value of $\varsigma_{I}$~(which can vary independently of $\fd_{\Dp}^{2}$ so that the above-mentioned limit for $(\fd_{\Dp}^{2}-4\varsigma_{I})$ can be met). The previous analyses thus do \emph{not} constrain the \emph{actual} CPT-violating neutrino LV scale $\Elvnu$ which, in D-foam interpretation of the effect~(\ref{eq:nsv}), is related to $\Md/\fd_{\Dp}^{2}$, and thus can only be limited via looking into energy-dependent TOF lags of neutrinos and the corresponding $\gamma$-rays, as in~(\ref{eq:cdf}).\vspace{0.05cm}

We note, finally, that the peculiarity of the foam in being \emph{transparent} to charged particle excitations~(as a result of charge conservation requirements~\cite{Ellis:2003ua,Ellis:2003if} discussed earlier) precludes one from translating the aforementioned results for the electron/positron sector into similar sensitivities to LV for neutrinos. Indeed, the different behavior of neutrinos from charged leptons in D-particle string scenarios implies a background-induced breaking of the SU(2) gauge symmetry of the SM. Otherwise, as illustrated in~\cite{Crivellin:2020oov} for~(field theory) models where the ordinary SU(2)$_{\textrm{L}}$ gauge invariance is still kept, the sizable Lorentz violation in neutrinos~(\ref{eq:nsv}) would clash with the previous LV bounds placed for electrons. For reasons just stated, D-foam models naturally avoid such conflicts~\cite{Li:2023wlo,Li:2022sgs}.
\ei
Before closing, we remark, for completeness, that UHE cosmic rays~(UHECRs) can cast tight constraints on LV~(primarily within the hadronic sector)~\cite{Aloisio:2000cm,Alfaro:2002ya,Jacobson:2002hd,Mattingly:2008pw,Galaverni:2008yj,Galaverni:2007tq,Maccione:2008iw,Maccione:2009ju,Scully:2008jp,Stecker:2009hj,Gonzalez-Mestres:2009evz,He:2024ljr,He:2025cfz}, as they possess the highest energies ever observed. In fact, much activity in this area was initially spurred by the earlier observational indication from the Akeno Giant Air Shower Array~(AGASA)~\cite{Takeda:1998ps} that the Greisen--Zatsepin--Kuz'min~(GZK) cutoff~\cite{Greisen:1966jv,Zatsepin:1966jv} might be removed for UHECRs striking CMB photons. The lack of sources for the AGASA post-GZK events $\gtrsim\Or(100~\textrm{EeV})$ leads to the so-called ``GZK paradox'', which may be explained via exotic physics, including LV modification of the cutoff feature. However, the anomaly has disappeared subsequently, due to improved sensitivity of experiments, especially High Resolution Fly's Eye~(HiRes) and Auger ones~\cite{HiRes:2007lra,PierreAuger:2007hjd,PierreAuger:2010gfm}, to such GZK~(proton) features. The existence of the photomeson production through the mechanism, $p+\gamma_{\textrm{CMB}}\rightarrow\D^{+}\rightarrow p+\pi^{0}$ or $n+\pi^{+}$, in turn severely constrains LV effects for protons~\cite{Xiao:2008yu,He:2025cfz} and/or pions, e.g.,~\cite{He:2024ljr}. In particular, LV threshold anomalies for $\pi^{0}$-production off the proton, similar to the photon--photon annihilation discussed earlier in the way that the kinematical threshold is modified, have been discussed in~\cite{He:2024ljr} and used, recently, in~\cite{He:2025cfz}, to set a two-sided limit on the proton LV, using also the current observational lack of proton decay~($p\rightarrow p\gamma$). However, one cannot use such results to set limits on D-particle string foam models, as in such cases, neither the incident charged high-energy proton $p$, nor the pion, whose constituents are also charged, interacts with the foam; the GZK effect is thus unaffected.

Similar stringent constraints from the nonobservation of a substantial fraction of UHE photons from the ensuing pion decay $\pi^{0}\rightarrow\gamma$~\cite{Maccione:2010sv,Maccione:2008iw,Maccione:2009ju,Galaverni:2008yj,Galaverni:2007tq}~(see also~\cite{Liberati:2011bp}) in the UHECR flux above 10~EeV were also obtained. Again, these results can be naturally evaded~\cite{Ellis:2010he,Mavromatos:2010pk}, in D-foam framework, at least as long as one sticks to the stretched-string formulation of the foam~\cite{Ellis:2008gg}, where the pertinent Lorentz-violating delays in photon propagation, associated with stringy uncertainties, do \emph{not} entail any modification of the local dispersion relations of photons. So that, those analyses, starting from assuming the photon MDRs induced at the level of a low-energy EFT formulation of QG-foam effects, are simply not applicable in the string/D-defect foam case. In such models, moreover, it may happen~\cite{Ellis:2010he} that as we noted in Section~\ref{sec:4}, by embedding the D-particle foam into low-scale string~(large extra-dimension) framework~\cite{Antoniadis:1999rm,Antoniadis:2000vd} with low $M_{s}$ of order $(10-100)$~EeV, i.e., the conventional GZK-cutoff scale, there would be no photons with energies higher than this that could not be completely absorbed by the defect. This serves as an explanation for their absence, in accordance with the observed lack of GZK-induced photons, thus providing additional reasons for evading the strong constraints of~\cite{Maccione:2010sv}.

\section{Conclusions}\label{sec:7}

While, presently, searches of potential violation of Lorentz and CPT invariance all lead  to inconclusive results, some of them have provided a clue of the existence of LV for cosmic photons and neutrinos~\cite{Shao:2009bv,Zhang:2014wpb,Xu:2016zxi,Xu:2016zsa,Amelino-Camelia:2017zva,Xu:2018ien,Liu:2018qrg,Chen:2019avc,Zhu:2021pml,Zhu:2021wtw,Zhu:2022usw,Song:2024and,Song:2025qej,Song:2025apj,Song:2025akr,Li:2020uef,He:2022gyk,Amelino-Camelia:2016ohi,Huang:2018ham,Huang:2019etr,Huang:2022xto}. As we have seen, Lorentz violation might play a role in the arrival delays of the more energetic photons emitted from GRBs and AGNs, as detected by Fermi, MAGIC, LHAASO, etc. A light-speed variation, which is subluminal and linearly depends on the photon energy, has been found from analyses combining data available at current stage, with a LV scale approaching the Planck energy, $\Elvg\sim 3\times 10^{17}$~GeV. This type of analysis also exposes an analogous effect for some of the IceCube neutrinos which are likely to be GRB neutrinos. The determined neutrino's LV scale, $\Elvnu$~(c.f.~(\ref{eq:nsv})), is not the same as that of photons but has the same magnitude. Previous observation shows that the neutrino speed can be both slower and faster than the constant light speed, implying CPT violation in the neutrino sector, while the signal of subluminal propagation was enhanced recently~\cite{Amelino-Camelia:2022pja}.

In this review we have discussed a class of Lorentz-violating quantum space-time foam models inspired from~(modern version of) string theory, involving brane-world scenarios, as a consistent framework that can explain the above results, in a natural fashion, following our recent studies in this regard~\cite{Li:2021gah,Li:2021eza,Li:2022sgs,Li:2023wlo,Li:2024crc}. The key point in the approach is the existence of lower-dimensional background brane defects, whose topologically nontrivial interactions with photons or at most electrically neutral particles and no charged ones, such as electrons, to which the D-brane foam looks transparent, induce the nontrivial refractive indices of the corresponding string vacuum. The latter are found proportional to the particle energy, thus being minimally suppressed by the effective QG mass, which is a complex function of many microscopic parameters in the model and is not simply given by the string mass scale, $\Ms$. Other unique properties or requirements which are indeed respected by D-foam models so that the findings~(\ref{eq:slsv}), (\ref{eq:nsv}) can be attributed to propagation in such a foam medium without clashing with other restrictions, include:~(i) photon stability against self-decay,~(ii) absence of birefringence, and~(iii) for certain stochastic foam, an energy violation that could prevent superluminal antineutrinos from decaying away.

The fact that, viable interpretations of the energy-dependent speed variations of cosmic photons and neutrinos are provided by string foam models within reasonable ranges of the model parameters, consistently with all other stringent tests of Lorentz invariance~(at least within the currently available sets of astrophysical data) is rather encouraging, as it provides one of the examples of experimentally finding signatures of QG. An unambiguous particle energy to time-of-arrival correlation linearly suppressed in the Planck mass, coupled with the observed lack of birefringence and many other LV reaction anomalies at the same order, might constitute, in our opinion, ``smoking-gun'' evidence of the models~\cite{Ellis:1999sf,Ellis:2000sf,Ellis:2004ay,Ellis:2005ib,Mavromatos:2007xe,Ellis:2008gg,Ellis:2009vq,Mavromatos:2010pk,Li:2021gah,Li:2021eza,Li:2022sgs,Li:2023wlo,Li:2024crc,Li:2009tt}, as other tight constraints forbid such a construction in EFT~\cite{Myers:2003fd}. Needless to say, all these findings need stronger confirmation by future experiments like, especially LHAASO, where photons from diverse sources will be observed. Still, if we are fortunate, results that may be established from a plethora of further studies may offer us opportunities to better defend the intriguing findings of the type endorsed here, and to restrict the classes of space-time theories. With all the progress to be made in the next decade or two, Lorentz violation and the phenomenology of D-foam will be bound, hopefully, to enter a more mature phase, and more insightful understanding of the riddles of our Universe can reasonably be hoped for in the not-so-distant future.

\vspace{6pt}



\authorcontributions{Both C.L. and B.-Q.M. contributed extensively to this project. All authors have read and agreed to the published version of the manuscript.}

\funding{This work was supported by the National Natural Science Foundation of China, grants number 12335006 and number 12075003. C.L. was supported in part by a \emph{Boya} Fellowship provided by Peking University and by China Postdoctoral Science Foundation~(CPSF), grant number 2024M750046. The work of C.L. was also financed by Postdoctoral Fellowship Program~(Grade B) of CPSF, grant number GZB20230032. C.L.'s work on this project also falls in the scope of the COST Action CA23130 Bridging high and low energies in search of quantum gravity~(BridgeQG).}

\institutionalreview{Not applicable.}

\informedconsent{Not applicable.}


\dataavailability{No new data were created or analyzed in this study. Data sharing is not applicable to this article.}


\conflictsofinterest{The authors declare no conflicts of interest.}



\abbreviations{Abbreviations}{
The following abbreviations are used in this manuscript:\\

\noindent 
\begin{tabular}{@{}ll}
AGASA & Akeno Giant Air Shower Array\\
AGN & Active galactic nucleus\\
CFT & Conformal field theory\\
CMB & Cosmic microwave background\\
CPT & Charge--Parity--Time\\
CPTV & CPT violation\\
EBL & Extragalactic background bight\\
EFT & Effective field theory\\
FGST & Fermi $\gamma$-ray Space Telescope\\
GBM & Gamma-ray Burst Monitor\\
GR & General relativity\\
GRB & Gamma-ray burst\\
GZK & Greisen--Zatsepin--Kuz'min\\
HAWC & High Altitude Water \v Cerenkov\\
HEGRA & High-Energy-Gamma-Ray Astronomy\\
H.E.S.S. & High Energy Stereoscopic System\\
HiRes & High Resolution Fly's Eye\\
IC & Inverse Compton\\
KM2A & Kilometer Square Array\\
KM3NeT & Cubic Kilometre Neutrino Telescope\\
LAT & Large Area Telescope\\
$\Lambda$CDM & Lambda~(Cosmological-constant) Cold Dark Matter\\
LHAASO & Large High Altitude Air Shower Observatory\\
LV  & Lorentz violation\\
MAGIC & Major Atmospheric Gamma-ray Imaging \v Cerenkov\\
MDR & Modified dispersion relation\\
QED & Quantum electrodynamics\\
QG & Quantum gravity\\
SM & Standard model\\
SME & Standard-model extension\\
SNR & Supernova remnant\\
SR & Special relativity\\
SSC & Synchro-self-Compton\\
TOF & Time-of-flight\\
TRIDENT & Tropical Deep-sea Neutrino Telescope\\
UHE & Ultra-high energy\\
UHECR & Ultra-high energy cosmic ray\\
UV & Ultraviolet\\
WCDA & Water \v Cerenkov Detector Array\\
WFCTA & Wide Field-of-view \v Cerenkov Telescope Array
\end{tabular}}




\begin{adjustwidth}{-\extralength}{0cm}

\reftitle{References}

\end{adjustwidth}
\end{document}